# Assessment of the probability of microbial contamination for sample return from Martian moons II: The fate of microbes on Martian moons

Kosuke Kurosawa[a], Hidenori Genda[b], Ryuki Hyodo[b], Akihiko Yamagishi[c], Takashi Mikouchi[d], Takafumi Niihara[e], Shingo Matsuyama[f] and Kazuhisa Fujita[g]

[a]Planetary Exploration Research Center, Chiba Institute of Technology, 2-17-1, Narashino, Tsudanuma, Chiba 275-0016, Japan.

[b]Earth–Life Science Institute, Tokyo Institute of Technology, 2-12-1 Ookayama, Meguro-ku, Tokyo 152-8550, Japan.

[c]Department of Applied Life Sciences, School of Life Sciences, Tokyo University of Pharmacy and Life Sciences, 1432-1, Horinouchi, Hachioji, Tokyo 192-0392, Japan.

[d]The University Museum, The University of Tokyo, 7-3-1 Hongo, Bunkyo-ku, Tokyo 113-0033, Japan.

[e]Department of Systems Innovation, School of Engineering, The University of Tokyo, 7-3-1 Hongo, Bunkyo-ku, Tokyo 113-8656, Japan.

[f]Aeronautical Technology Directorate, Japan Aerospace Exploration Agency, 7-44-1, Jindaijihigasi-machi, Chofu, Tokyo 182-8522, Japan.

[g]Institute of Space and Astronomical Science, Japan Aerospace Exploration Agency, 3-1-1, Yoshinodai, Chuo-ku, Sagamihara, Kanagawa 252-5210, Japan.

*Corresponding author
Kosuke Kurosawa, Ph. D
Planetary Exploration Research Center, Chiba Institute of Technology
E-mail: kosuke.kurosawa@perc.it-chiba.ac.jp
Tel: +81-47-4782-0320; Fax: +81-47-4782-0372; ORCID ID: 0000-0003-4965-4585

**Abstract**

This paper presents a case study of microbe transportation in the Mars-satellites system. We examined the spatial distribution of potential impact-transported microbes on the Martian moons using impact physics by following a companion study (Fujita et al., accepted). We used sterilization data from the precede studies (Patel et al., 2018; Summers, 2017). We considered that the microbes came mainly from the Zunil crater on Mars, which was formed during 1.0–0.1 Ma. We found that 70–80% of the microbes are likely to be dispersed all over the moon surface and are rapidly sterilized due to solar and galactic cosmic radiation except for those microbes within a thick ejecta deposit produced by natural meteoroids. The other 20–30% might be shielded from radiation by thick regolith layers that formed at collapsed layers in craters produced by Mars rock impacts. The total number of potentially surviving microbes at the thick ejecta deposits is estimated to be 3–4 orders of magnitude lower than at the Mars rock craters. The microbe concentration is irregular in the horizontal direction due to Mars rock bombardment and is largely depth-dependent due to the radiation sterilization. The surviving fraction of transported microbes would be only ~1 ppm on Phobos and ~100 ppm on Deimos, suggesting that the transport processes and radiation severely affect microbe survival. The microbe sampling probability from the Martian moons was also investigated. We suggest that sample return missions from the Martian moons are classified into *Unrestricted Earth-Return* missions for 30 g samples and 10 cm depth sampling, even in our conservative scenario. We also conducted a full statistical analysis pertaining to sampling the regolith of Phobos to include the effects of uncertainties in input parameters on the sampling probability. The most likely probability of microbial contamination for return samples is estimated to be two orders of magnitude lower than the $10^{-6}$ criterion defined by the planetary protection policy of the Committee on Space Research (COSPAR).

**1. Introduction**

Microbe transportation between planetary bodies due to impact ejection, which is often referred to as the "(Litho-)Panspermia", is widely thought to be important in the context of the origin of life (e.g., Horneck et al., 2008; Melosh, 1988; Price et al., 2013). In recent years, an extraterrestrial planetary system in which multiple planets are present in the habitable zone around TRAPPIST-1 has been discovered (Gillon et al., 2016; 2017), leading to further consideration of the importance of impact-driven microbe transportation (Krijt et al., 2017; Lingam and Loeb, 2017). Martian and lunar meteorites found on Earth represent direct evidence for material transportation between planetary bodies due to impact ejection (e.g., Bogard and Johnson, 1983; Eugster, 1989; Hill et al., 1991; Hidaka et al., 2017; Marvin, 1983; Nyquist et al., 2001; Wiens and Pepin, 1988;).

Recently, the Japan Aerospace eXploration Agency (JAXA) announced that the next target for a sample return mission is one of the Martian moons (i.e., Phobos or Deimos). The presence of Martian meteorites on Earth implies that Mars rocks have been transported to its moons throughout history (e.g., Chappaz et al., 2013; Melosh, 2011; Ramsley and Head, 2013). Given the possibility that organisms exist on Mars, microbial-bearing rocks might have been transported from Mars to its moons. Consequently, the transport processes in the Martian satellite system are significant for both astrobiology and planetary protection (Chappaz et al., 2013; Fujita et al., in submission; Melosh, 2011, Patel et al., 2018; Summers, 2017).

In this study, we extend the work of our companion study (hereafter referred to as Paper I) (Fujita et al., accepted) and investigate the fate of impact-transported microbes on the Martian moons using our knowledge of impact-related processes and a radiation-induced sterilization model in the space environment at a Martian orbit. According to the COSPAR Planetary Protection Policy (PPP), the probability that a single unsterilized particle of ≥10 nm in diameter is in a sample returned from extraterrestrial bodies should be <$10^{-6}$ for *Unrestricted Earth-Return* missions. This criteria is termed as REQ-10. We determined the microbial contamination probability to assess whether the criteria of REQ-10 would be met for a future sample return mission from Martian moons.

Here, we mention about the guiding principles employed in this study:

(1) We considered the planetary protection issue about the sample return mission from the Martian moons by following the REQ-10 criterion. We do not discuss the pros and cons of the REQ-10 in this study.
(2) We obtained a quantitative and conservative value pertaining to microbial contamination probability from the return sample based on our current best knowledge. In other words, additional experimental works are beyond the scope of this study.

(3) In the case where scientific judgements could not be made, we employed assumptions, which yield higher values of the microbial contamination probabilities, to obtain conservative estimates.

In Paper I, the probability density function (PDF) of the density of the potential microbes in the Martian environment was derived, based on microbe densities at terrestrial analog sites. The survival rate of the potential microbes after launch and atmospheric passage into space was then estimated to be ~0.1. In this study, we used this result as an input parameter, and provide realistic constraints on the fate of microbes after transportation to a Martian moon.

The remainder of this manuscript is organized as follows. Section 2 describes the supporting data regarding sterilization taken from previous studies (Sections 2.1 and 2.2). We used the recent sterilization data for various types of microbe obtained from the research group named "Sterilisation Limits for Sample Return Planetary Protection Measures (SterLim)" (Patel et al., 2018; Summers, 2017). This group has extensively studied the sterilization processes of microbes in different types of experiment, including hypervelocity impacts and radiation. We describe the potential number of microbes transported from Mars to its moons determined by Paper I and other related studies (Genda et al., in preparation; Hyodo et al., in preparation) in Section 2.3. We then investigate the spatial distribution of the transported Mars rocks on the moons in Section 3, in which we consider crater formation (Section 3.1), dust torus production (Section 3.2), and the role of background meteoroid impacts (Section 3.3). We present the temporal changes in the number of surviving microbes in Section 4.1 and the sampling probability of the microbes from each moon in Section 4.2. The propagation of uncertainties is discussed in Section 4.3. We show that the microbe concentration on the present Martian moons is expected to be irregular in the horizontal direction due to stochastic impacts and heterogeneous in the vertical direction due to the depth-dependent time constant pertaining to the radiation-induced sterilization. There are three types of location, depending on the timescale of the radiation-induced sterilization: (1) a thin global microbial layer (~0.1 mm thick); (2) the global layer, but covered by a thick ejecta deposit produced by impacts of natural meteoroids (~0.1 mm thick, but covered by an ejecta deposit with a thickness of >3 mm); and (3) the interiors of craters produced by the impact of Mars rocks (~1 m thick). Figure 1 shows a schematic diagram of the various processes taking place on the Martian moons that were considered in this study. The details of these processes are presented in Sections 3 and 4. Finally, we summarize the outcomes of this study in Section 5.

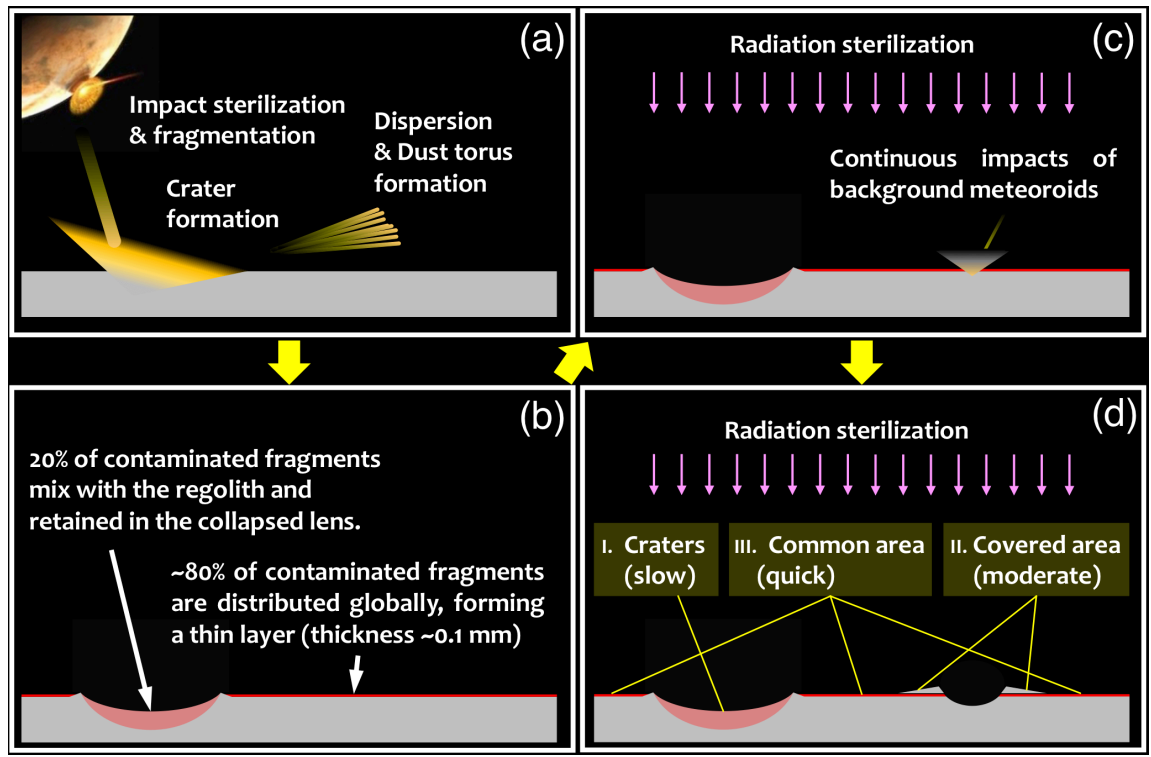

**Figure 1.** Schematic diagram of the processes considered on the Martian moons (see Sections 3 and 4). The time intervals proceed from (a) to (d). In (d), three distinctive locations with different timescales for radiation-induced sterilization are shown.

## 2. Supporting data and initial conditions

### 2.1. Impact-induced sterilization

In hypervelocity impacts of Mars rocks onto the surface of its moons, the incident objects have a high kinetic energy, and considerable shock-heating occurs in the incident materials on impact. As the impact velocity increases, the degree of shock-heating would become more significant in the incident materials. This suggests that impact sterilization would be greater at higher impact velocity. We used the experimental data from the SterLim impact test (Patel et al., 2018) to assess this further. The survival rates of various types of microbe after hypervelocity impacts at velocities of 0.5 to 2.2 km/s were reported by Patel et al. (2018) and are summarized in Fig. 2. Patel et al. (2018) used a basalt cylinder of 3 mm in diameter as a projectile and a regolith simulant for the Martian moons as a target. The microbes were loaded into a hole that was 0.75 mm in diameter, which was drilled into the center of a flat plane of the basalt cylinder.

More than 10 previous studies have investigated the degree of sterilization during impact shocks (compiled by Barney et al., 2016) prior to the SterLim study. Unfortunately, the results of these previous studies are inconsistent when the survival rates after the shocks are related to the peak pressures experienced in the experiments (Barney et al., 2016). This might reflect the different experimental procedures used in the previous studies, such as single shock and reverberation. In addition, the peak pressure is not an ideal variable to correlate with sterilization, as temperature (or entropy) is expected to be the more important parameter because it determines whether the molecular systems supporting the activities of individual cells are degraded.

Burchell et al. (2004) and Price et al. (2013) accelerated microbe-bearing projectiles into water or agar targets. In contrast, Willis et al. (2006), Horneck et al. (2008), Hazell et al. (2010), and Hazeal et al. (2014; 2017) accelerated a flyer plate onto microbes between metal or rock layers. The former and latter experiments lead to single shock compression and shock reverberation, respectively. The pressure–temperature path in a reverberation-type experiment is very different from that in natural impact events (Horneck et al., 2008; Ivanov and Deutsch, 2002), although reverberation-type experiments allow collection of the shocked microbes with a high recovery rate. Consequently, it is appropriate to use the results from single shock-type experiments. Burchell et al. (2001; 2003) and Fajardo-Cavazos et al. (2009) also conducted single shock-type experiments; however, microbes were embedded in the target in their experiments. In this case, microbes might experience compression and thermal pulsing due to the impacts, depending on their initial location in the target. This does not correspond to the situation considered in the present study, whereby microbe-bearing Mars rocks collide with a moon surface.

Thus, we now discuss the differences between the SterLim impact tests and the studies of Burchell et al. (2004) and Price et al. (2013), who used water and agar targets. The shock impedance of agar is approximated by that of water (Price et al., 2013). In contrast, the SterLim impact tests were designed to simulate the collision between Mars rocks and the moon surface as closely as possible. There is a fundamental difference between the target being water and rocky material, as the specific heat of the two materials results in a higher shock temperature of the rocky material. Thus, the temperatures experienced in the SterLim impact tests would be much higher than those of the previous studies. In fact, Burchell et al. (2004) and Price et al. (2013) were able to detect trace amounts of surviving microbes even at impacts of >5 km/s, possibly due to the low shock temperature.

Given these considerations, we only use the SterLim data in this study. Kurosawa and Genda (2018) reported that the degree of impact heating is much higher than previously expected owing to plastic deformation in rocky materials, especially in the case of low-velocity impacts (<10 km/s). This heat source has been overlooked for a long time (Melosh and Ivanov, 2018) but the effects were naturally included in the SterLim impact tests. The findings of Kurosawa and Genda (2018) support the validity of our data selection.

**2.2. Improved model for impact sterilization**

It is clear that the microbial survival rate must decrease as the impact velocity increases, as more kinetic energy results in greater shock-heating. Such a trend is clearly evident from the impact test results of the SterLim study (Fig. 2). However, in the SterLim study, the microbial survival rate was assumed to be uniquely 0.1, regardless of the impact velocity. This is because the impact velocities of the Mars ejecta considered in the SterLim study were <2 km/s on average. However, our trajectory analysis predicts that a major proportion of Mars ejecta should have impact velocities of >2 km/s (Hyodo et al., in preparation). Consequently, we developed a new physical model based on the SterLim data to estimate the microbial survival rate at impacts of >2 km/s, as follows.

The microbial survival rate for hypervelocity impacts of Mars ejecta on the surface of the Martian moons can be reasonably defined as follows:

$$N/N_0 = \exp(-CV^\alpha) \qquad (1)$$

where $N$ and $N_0$ are the final and initial number of microbes, $V$ is the impact velocity of the Mars ejecta, and $C$ and $\alpha$ are constants (see below). The above definition must be physically acceptable since the kinetic energy of the incident object per unit mass is proportional to $V^2$, and so is the temperature increase in the fragments after the impact. If the transfer from kinetic to thermal energy is ideal, then $\alpha = 2$. However, a proportion of the kinetic energy might be transferred to irreversible compression and the internal degrees of freedom of the target materials, which dissipates without the rise in temperature of the incident materials. For these reasons, $\alpha$ is not >2, but close to 2.

The variable parameters were adjusted so that the impact test data from the SterLim study were satisfactorily reproduced. However, direct least-squares fitting to the SterLim data ($\ln N/N_0$) yielded an unrealistic value of $\alpha = 4.3$ as the optimized value. This is probably because the velocity range is limited, whereas the data are widely scattered. For this reason, to determine $\alpha$ we assumed that only 20% of the kinetic energy is transferred to the temperature increase for sterilization at $V = 2.0$ km/s. Thus, $V^\alpha = 0.2V^2$ and $V = 2,000$ yields $\alpha = 1.8$. Least-squares fitting was then applied to $\ln N/N_0$ with $\alpha = 1.8$. In this procedure, the data points where $\ln N/N_0 < -10$ for $V^{1.8} > 5 \times 10^5$ were excluded from the fit, because these data points are close to the detection limit and have a low signal-to-noise ratio. Exclusion of these data allows more conservative estimation of the microbial survival rate, since the low microbial survival rate data are not taken into account. The final fitting result is illustrated in Fig. 2a, which yielded $C = 9.5 \times 10^{-6}$.

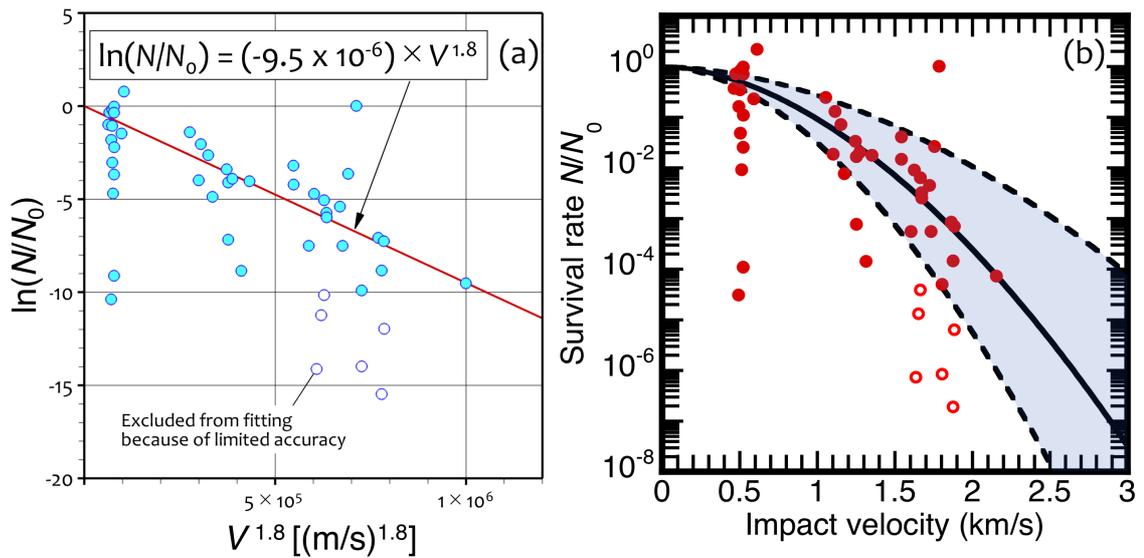

**Figure 2.** (a) Microbial survival rate during impacts as a function of impact velocity, and a least-squares regression of the impact test data from the SterLim study ($\ln N/N_0$). (b) The fitting line with the uncertainty bound (blue shaded region) by Eq. (3). Some data points were excluded from the least-squares regression due to their low signal-to-noise ratio (open symbols).

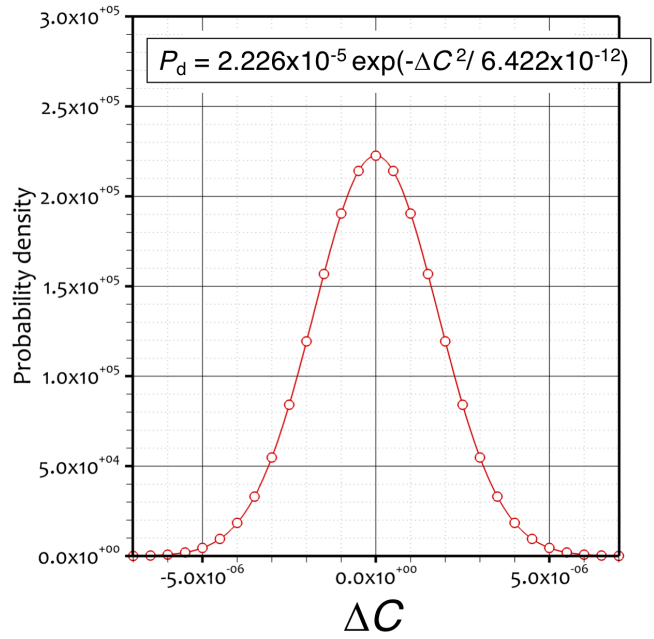

**Figure 3.** Variation of the determination coefficient from a least-squares regression of the impact test data from the SterLim study ($\ln N/N_0$).

A probabilistic function of the microbial survival rate was deduced from the determination coefficient of the least-squares fit. Variation of the determination coefficient for $\Delta C = C - 9.5 \times 10^{-6}$ is shown in Fig. 3. Based on the difference of the determination coefficient from the peak value, the probabilistic function was determined as follows:

$$P_{d,i} = 2.226 \times 10^5 \exp\left(-\Delta C/6.422 \times 10^{-12}\right) \tag{2}$$

where the standard deviation is $\sigma = 2.226 \times 10^5$. The determined probabilistic function is plotted in Fig. 3. The final form of the microbial survival rate obtained in this study is given by

$$N/N_0 = \exp\left[-(9.5 \pm 4.3) \times 10^{-6} \, V^{1.8}\right] \tag{3}$$

where $C = 9.5 \times 10^{-6}$ is the fiducial case and the uncertainty corresponds to 2.5 times the standard deviation. Figure 2b shows the fitting line with the 2.5σ uncertainty bounds by Eq. (3). The new model well reproduces the trend in the survival rate with increasing impact velocity although there is a large data scatter. The probabilistic function of Eq. (2) was used for the statistical analysis described in Section 4.3. The verification of the model accuracy is not sufficient for impact velocities above 2.0 km/s, and it is strongly recommended that further impact tests in the higher velocity range are conducted in the future. It should be mentioned here that it is impossible to determine which exponents ($\alpha = 1.8$ or $\alpha = 4.3$) are accurate by scientific discussion without further data with a high accuracy and a good reproducibility. Nevertheless, we employed $\alpha = 1.8$ by following the guiding principle mentioned in Section 1 because it makes a conservative estimate. Even if we employed $\alpha = 4.3$, the main conclusion of this study does not largely change because it yields a smaller microbial contamination probability than the value in the case with $\alpha = 1.8$.

## 2.3. Radiation-induced sterilization

Microorganisms transported to the Martian moons are expected to be sterilized over time by solar and galactic cosmic radiation (γ-rays, protons, and α particles). Given the SterLim study provided comprehensive data for radiation sterilization based on thorough radiation inactivation tests (Patel et al., 2017), these data are incorporated in this study without modification. The time constant ($TC$), which is the time required to sterilize microorganisms by 1/e, was also taken from the SterLim study (Summers, 2017; Patel et al., 2017). The microbial survival rate is calculated as follows:

$$\frac{N}{N_0} = \exp\left(-\frac{t}{TC}\right), \tag{4}$$

where $N_0$ and $N$ are the microbial density at the beginning and after time $t$, respectively, and $t$ is the time in years. The constant $TC$ depends on the depth where microorganisms are deposited.

As the time increases and depth decreases, the microbial survival rate quickly decreases. Given the regolith layer on the moon surface acts as a shield against radiation, the sterilization rate depends strongly on the depth below the surface. The required time for sterilization $t_{req}(H)$ at a given depth $H$ is given by

$$t_{req}(H) = TC \ln\left(\frac{N_0}{N_{th}}\right), \tag{5}$$

where $TC$, $N_{th}$, and $N_0$ are the depth-dependent time constant for sterilization taken from the SterLim report, and the threshold and initial microbe density, respectively. Figure 4 shows $t_{req}$ as a function of depth. In this calculation, $N_{th}$ and $N_0$ were set to $10^{-5}$ and $10^7$ cells/kg, respectively. The value of $N_{th}$ was chosen to satisfy the REQ-10 criterion in the case of a 100 g sample. The initial microbe density $N_0$ was set to 0.1 times the fiducial value (see Paper I) of the potential microbe density on the Martian surface. A factor of 0.1 was used to express the impact sterilization during the ejection from Mars as discussed in Paper I. We used the $TC$ values of MS-2 coliphage, which exhibited the most radiation resistance in the SterLim radiation tests, as a conservative scenario throughout this study. The radiation resistance of the MS-2 coliphage may be close to the theoretical upper limit pertaining to the "Earth-type" life, which is supported by DNA/RNA having the genetic information. The MS-2 coliphage has been known as the smallest virus (~27 nm (Straus and Sinsheimer, 1963)) having only 3,569 nucleotides (Fiers et al., 1976). They encode just four proteins (van Duin and Tsareva, 2006), which are required for sustaining their activity as a virus, including the infection to the bacterium *Escherichia coli*. Because of its smallness, the MS-2 coliphage exhibits the highest radiation resistance in the known microbes. Consequently, we have considered that the MS-2 coliphage was one of the best choices to obtain a conservative estimate pertaining to the microbial contamination probability. Figure 4 clearly shows that the microbes are sterilized within 2 Myr down to 10 cm below the surface. The mission requirements pertaining to the MMX mission are that the sampling system collects more than 10 g sample during one sampling operation and collects samples from the surface layer to the depth deeper than 2 cm (Kawakatsu et al., 2017). This depth is desirable to recover "fresh samples" that are not altered by space weathering due to cosmic ray exposure. If microbes from Mars are located in the uppermost layer of the moons, then they are rapidly sterilized within 2 kyr.

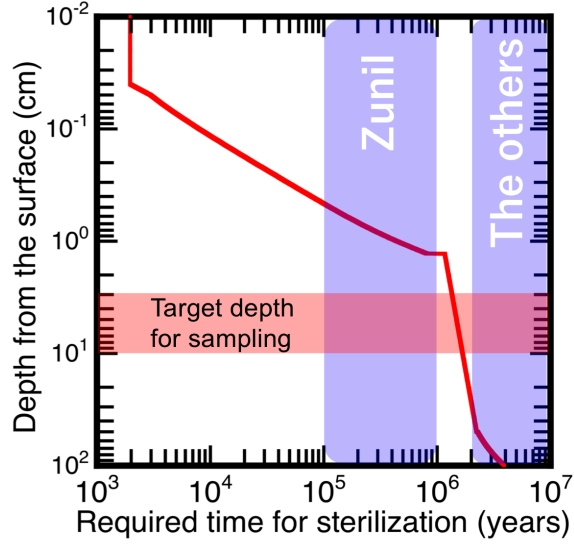

**Figure 4.** Time required for sterilization by radiation $t_{req}$ (Eq. (5)) as a function of depth. The age estimates of the known craters on Mars and target depth for sample collection currently selected by the JAXA MMX mission are also shown as the shaded regions. If $t_{req}$ is multiplied by a factor of 0.036, the time constant $TC$ obtained by the SterLim radiation tests can be reproduced.

Given that the samples of the Martian moons will be collected by boring, it is more appropriate to calculate the microbial survival rate average for the specific depth by integrating the microbial survival rate over a depth interval and dividing by depth. The average radiation survival rate $\eta(t, H)$ integrated down to a given depth $H$ is given by

$$\eta(t, H) = \frac{\int_0^H \exp\left(-\frac{t}{TC(h)}\right) dh}{H}, \qquad (6)$$

where $t$ is the time after transportation. Figure 5 shows $\eta(t, H)$ as a function of the depth from the surface at various times. In general, the average microbial survival rate is lower than the local survival rate at each depth, because microbial density decreases toward the exposed surface.

**2.4. Microbe survival immediately before arrival**

In this section, we briefly summarize the results of Paper I and the other companion papers (Genda et al., in preparation; Hyodo et al., in preparation). In Paper I, we derived the PDF of the potential microbe density in the Martian environment. It was assumed that a Gaussian of $\log_{10} n$ applies, where $n$ is the microbial density. The center value and standard deviation ($\sigma$) were assumed to be $-7.5$ and $0.5$, respectively. Hereafter, we use $n = 10^8$ cells/kg as a fiducial case, except in Section 4.4.

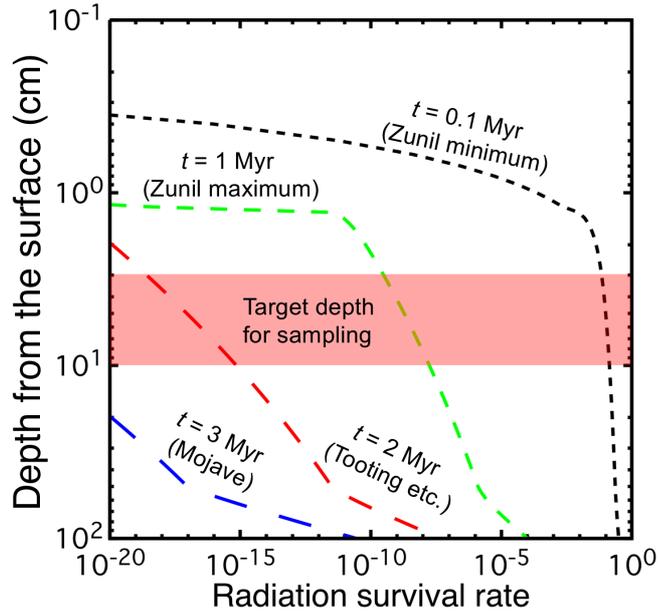

**Figure 5.** Average survival rate due to radiation as a function of depth. The results at four different times are shown. The times and crater names formed at these times are annotated next to the lines. As in Fig. 4, the target depth for sample collection currently used by the JAXA MMX mission are shown by shading.

In this study, we assume that only hypervelocity impacts on Mars can cause microbe transport. This is expected to be a reasonable assumption because endogenous processes cannot explain the acceleration of surface rocks on Mars to the escape velocity (e.g., De Carli et al., 2007). In contrast, the ejection process, which is termed 'impact spallation', can produce high-speed ejecta (e.g., Artemieva and Ivanov, 2004; Head et al., 2002; Kurosawa et al., 2018; Melosh, 1984; Vickery and Melosh, 1987). We investigated the high-speed material ejection during impacts using a three-dimensional (3D) smoothed particle hydrodynamics (SPH) code, and derived the mass–velocity distribution at various impact velocities and angles (Genda et al., in preparation). The 3D SPH code has been widely used in planetary science (Fukuzaki et al., 2010; Genda et al., 2015; Kurosawa et al., 2018) and was validated via a comparison with an impact experiment (Okamoto et al., in revision).

We focused on the Zunil crater (7.7°N, 166°E; 10.1 km in diameter) on Mars as the main source of the microbes potentially living on the current moons. Zunil is the youngest known ray crater with a diameter of >10 km on Mars. Ray systems around a host crater are considered one of the most prominent signatures of fresh craters on planetary bodies (e.g., Hawke et al., 2004; Preblich et al., 2007). A lower limit of the formation age of Zunil crater has been estimated to be 0.1 Ma, based on a crater chronology model (e.g., Hartmann et al., 2010). We also used another constraint on Zunil crater, which is the impact direction. The geology of Zunil and the area around the crater, including the spatial distribution of the ray system and the secondary craters, has been extensively studied (e.g., Tornabene et al., 2006; McEwen et al., 2005; Preblich et al., 2007; Hartmann et al., 2010). It has been constrained that Zunil crater was formed by a moderately oblique impact from the east-northeast (Preblich et al., 2007). Thus, the impact direction for Zunil crater in our model was chosen from the northeast to the east (Hyodo et al., in preparation).

In Section 2.3, we showed that the microbes are sterilized within 2 Myr down to a depth of 0.1 m from the surface. Thus, we neglected the contribution by the other larger craters to microbe transport. The formation ages of other well-known craters, such as Corinto (13.8 km), McMurdo (23 km), Tooting (29 km), and Mojave (58 km), have been estimated to be >2 Ma (Golombek et al., 2014; Hartmann et al., 2010; Malin et al., 2006; Werner et al., 2014). Although there are ~10 smaller, young ray craters on Mars with similar formation ages of 1.0–0.1 Ma (Hartmann et al., 2010; Tornabene et al., 2006), their contribution to microbe transport is likely to be negligible compared with the Zunil event because the transported mass from Mars to its moons is roughly proportional to the cube of crater diameter (Hyodo et al., in preparation). During impact ejection, Mars rocks are likely to experience compression and heating. We used a survival rate of 0.1 as a conservative value during ejection, as discussed in Paper I.

An ejection velocity of >3.8 km/s is required to reach the Phobos orbit in the Martian gravity field. The launched Mars rocks must interact with the Martian atmosphere. Thus, we conducted an aerodynamic analysis to assess the effects of aerodynamic heating on the survival rate of the launched microbes during atmospheric passage in Paper I. We confirmed that sterilization during the atmospheric passage can be neglected when the diameters of Mars rocks are greater than 0.1 m, which corresponds to a minimum diameter required for penetrating the Martian atmosphere (Artemimeva and Ivanov,

2004; Paper 1). Consequently, the survival rate due to aerodynamic heating was set to unity.

The travel of the ejected Mars rocks in cismartian space was analytically solved as a two-body problem (Hyodo et al., in preparation). The initial positions and velocity vectors of the Mars rocks were taken from the 3D SPH calculation. We obtained the PDF of the transported mass of Mars rocks by conducting 10,000 Monte Carlo runs for the Zunil-forming impact event (Fig. 6). The impactor distributions for the Zunil-forming impactor in terms of size, impact velocity $v_{imp}$, and impact angle $\theta_{imp}$ were taken into account in the Monte Carlo runs (Hyodo et al., in preparation). The phase angles of the Martian moons were randomly selected. The average values of the transported mass colliding with Phobos and Deimos $M_{p,total,ave}$ are $2.0 \times 10^6$ and $3.8 \times 10^4$ kg (Fig. 6), respectively. We used the average values as a fiducial case throughout this manuscript. Given that >80% of the runs yielded a smaller transported mass, the average values are considered to be a conservative estimate. Consequently, the total numbers of transported microbes from Mars to Phobos and Deimos were estimated to be $2 \times 10^{13}$ and $3.8 \times 10^{11}$ cells, respectively, in the fiducial case. Our trajectory analysis also provided the PDFs of $v_{imp}$ and $\theta_{imp}$ of the Mars rocks colliding with the moons (Fig. 7a and b). As for Zunil-forming impactor, since we have limited impact direction to those from east-northeast, the ejecta is produced to the west-northwest direction. On the other hand, the orbital directions of Phobos and Deimos are from west to east. Therefore, the ejecta experiences nearly head-on collisions with Phobos and Deimoms. Consequently, the impact velocities of the Mars rocks colliding with the moons tends to be higher than the orbital velocities of Phobos and Deimos (~2 km/s for Phobos and 1.4 km/s for Deimos). We used the $v_{imp}$ and $\theta_{imp}$ distributions to address impact-driven processes on the moons in Section 3.

## 3. Distribution of Mars ejecta fragments over the Martian moons

In this section we discuss impact bombardment of the microbe-bearing Mars rocks on each of the moons and their distribution after the bombardment. We assumed that all the Mars rocks are spheres of 0.1 m in diameter, because the size–frequency distribution (SFD) of the ejected Mars rocks is highly uncertain. This size corresponds to the minimum size required for penetrating the Martian atmosphere into space (e.g., Artemieva and Ivanov, 2004; Paper I). This assumption yields an upper limit of the cumulative surface area of the resultant impact craters on the regolith of each of the moons, because it is controlled by frequent impacts by small impactors. In contrast, the mixing depth of the microbes into the regolith would be a minimum under this assumption. Consequently, for the Zunil cratering event, the total numbers of impacts to Phobos and Deimos were estimated in the fiducial case to be $1.4 \times 10^6$ and $2.7 \times 10^4$, respectively, with a projectile density $\rho_p = 2.7 \times 10^3$ kg/m$^3$.

We used a Monte Carlo approach to take into account the effects of the $v_{imp}$ and $\theta_{imp}$ distributions on the outcomes of the Mars rock bombardment. The *Mersenne Twister* algorithm (Matsumoto and Nishimura, 1998) was used to produce the random numbers in the Monte Carlo calculations. We randomly assigned $v_{imp}$ and $\theta_{imp}$ to each microbe-bearing rock from the distributions (Fig. 7). This treatment allowed us to characterize each impact event. The impact outcomes with a given set of impact conditions can be predicted by using the results from previous studies in the field of impact physics (e.g., Daly and Schultz, 2016; Housen et al., 1983; Melosh, 1989; Schmidt and Housen, 1987).

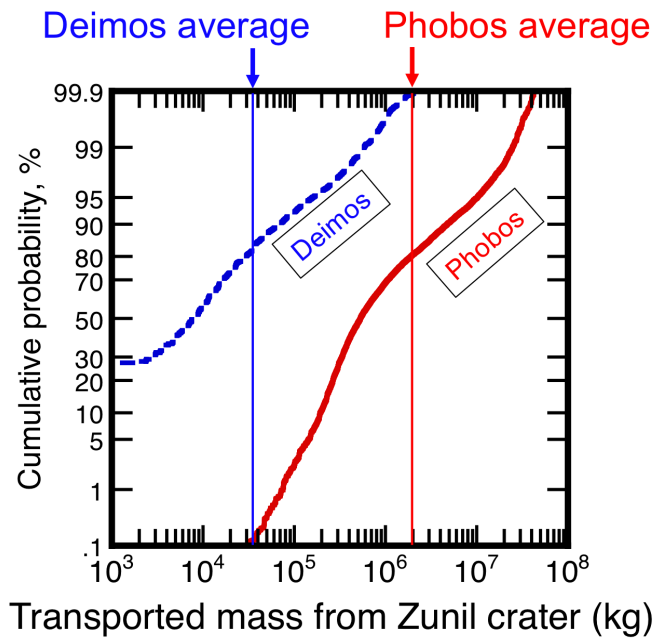

**Figure 6.** Cumulative probabilities of transported mass from Zunil crater obtained in a companion study (Hyodo et al., in preparation). The results for Phobos (red solid line) and Deimos (blue dashed line) are shown. The average values of the transported masses are shown as vertical lines.

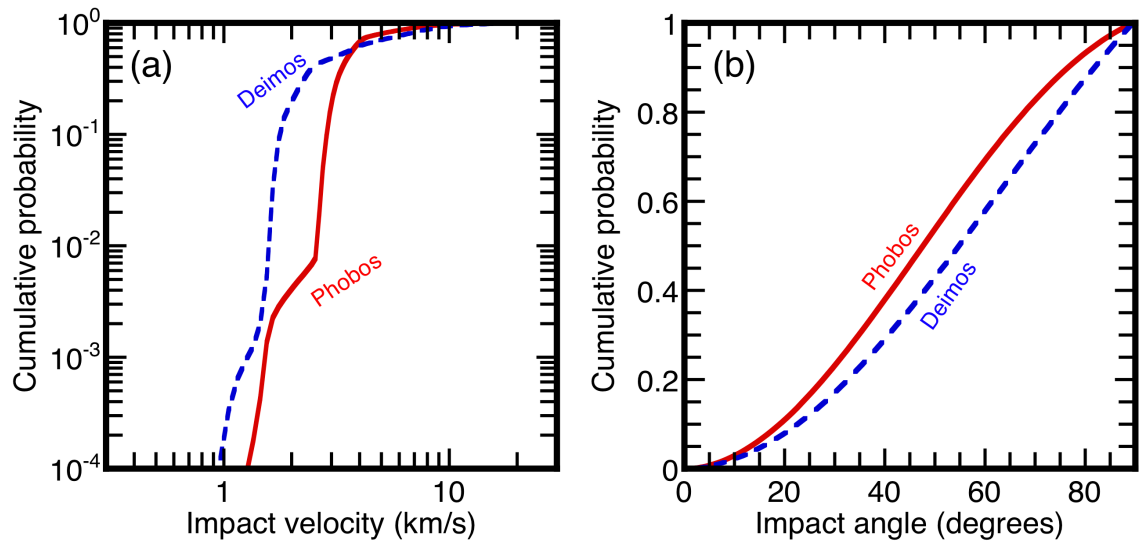

**Figure 7.** Distributions of the (a) impact velocity and (b) impact angle of Mars rocks colliding with Phobos (red solid lines) and Deimos (blue dashed lines).

### 3.1. Crater formation

Hypervelocity impacts of Mars rocks onto the surfaces of each moon should produce impact craters (e.g., Melosh, 1989). In general, impact-driven gardening process is rather complex because various effects, such as the fragmentation and the ejection of the bedrocks embedded beneath the regolith layer and secondary impacts of the large ejecta blocks onto the regolith layer, should be taken into account (e.g., Ramsley and Head, 2017). In this study, we assumed that all the impacts of Mars rocks occur on a regolith layer and that secondary impacts could be neglected, because each moon surface is globally covered by a regolith layer with a thickness of >20 m (e.g., Thomas, 1998), which is much thicker than the diameter of the Mars rocks (0.1 m). In addition, we assumed that the cratering processes are controlled by gravity rather than the strength of the regolith layer. A gravity-dominated regime is reasonable because granular materials generally have low strength. A somewhat higher sampling probability for the microbes from the Mars rock crater is expected if the cratering process on the moons is classified as a strength-dominated regime, because the resultant craters would be smaller than those in a gravity-dominated regime.

The longitudinal stress in a projectile penetrating into the regolith layer is likely to exceed the compressive strength of the microbe-bearing Mars rocks. For example, the compressive strengths of intact basaltic rocks on Earth have been reported to be 0.17–

0.48 GPa (e.g., Mizutani et al., 1990). In contrast, the dynamic ram pressure $P_{ram}$ ~ $\rho_t v_{imp}^2/2$ is estimated to be 1–10 GPa for impacts at 1–3 km/s, where $\rho_t$ ~ $2 \times 10^3$ kg/m$^3$ is the regolith density (e.g., Andert et al., 2010). Thus, the Mars rocks will be broken into fine fragments during penetration. Some of these fragments are likely to adhere to the wall of the growing crater (e.g., Elbeshausen et al., 2013) and mix with the regolith particles (e.g., Daly and Schultz, 2016; Ebert et al., 2014). After transient crater formation, the wall of the transient crater collapses due to gravity, resulting in granular flow towards the crater center. This is often referred to as the "modification stage" in the cratering process (e.g., Melosh, 1989; Barnouin-Jha et al., 2007). This granular flow would enhance the mixing between the microbe-bearing Mars rock fragments and regolith. Thus, we assumed that the Mars rock fragments were homogeneously mixed into the collapsed lens deposited on the final crater floor. During the collapse, the crater diameter and depth gradually increase and decrease, respectively, to satisfy mass conservation (e.g., Melosh, 1989). Such modification processes would occur in the weak gravity field of each of the moons, as well as under the gravity-dominated cratering regime, although the characteristic time for modification becomes somewhat longer than in the case of the planets (e.g., Mars) owing to the low gravity. Hereafter, the produced volume of material deposited on the crater floor is referred to as the "collapsed lens" and the resultant impact craters of the Mars rocks with the collapsed lenses are referred to as "Mars rock craters". The counterpart of the retained volume of the projectile would be ejected from the crater. The fate of this ejecta is discussed in Section 3.2.

To investigate the microbe concentration of the collapsed lens in a Mars rock crater, knowledge of the fraction of projectile retention $\psi$ is necessary. Unfortunately, the dynamics of projectile deformation and fragmentation during impact processes are not fully understood. In this study, we used the experimental data of Daly and Schultz (2016). Although the impact velocity was fixed at 4.5–5.0 km/s in their experiments, which is higher than the average impact velocity of the Mars rocks, this study provides a systematic dataset of $\psi$ at impact angles of 30° to 90° measured from the target surface. We derived an empirical formula by combining this dataset and two physical constraints, which are (1) $\psi = 0$ at an impact angle $\theta_{imp} = 0°$ and (2) $d\psi/d\theta_{imp} > 0$, as follows:

$$\psi = 0.718 - 1.01\cos\theta_{imp} + 0.294\cos^2\theta_{imp}. \qquad (7)$$

According to the orbital calculations (Hyodo et al., in preparation), the impact angles of Mars rocks onto Phobos and Deimos are likely to be nearly isotropic (Fig. 7b). By convoluting the $\theta_{imp}$ frequency distribution, the average retained fraction $\psi_{ave}$ of the Mars rocks in the final crater on each moon was estimated to be 22% for Phobos and 29% for Deimos.

The total mass of the collapsed lens $M_c$ in a Mars rock crater was roughly estimated by considering the geometries of the transient and final craters (Melosh, 1989, pp. 129). Three assumptions are made in calculating $M_c$: (1) both the transient and final crater have a parabolic shape; (2) the depth-to-diameter ratio of the final crater is 0.2; and (3) the depth of the transient crater $H_{tr}$ is the sum of the depth of the final crater $H_f$ and the thickness of the collapsed lens $H_c$. Figure 8 shows a schematic cross-section of the final crater considered here. Using these assumptions, we can estimate $M_c$ as follows:

$$M_c = \frac{\pi}{80} \rho_t D_f^3, \tag{8}$$

where $D_f$ is the diameter of the final crater. We estimated $D_f$ under a given set of impact conditions using the π-group scaling law and an empirical relationship linking the transient crater diameter $D_{tr}$ with $D_f$ (e.g., Melosh and Vickery, 1989) as follows:

$$D_f = 1.25 D_{tr} \tag{9}$$

and

$$D_{tr} = \left(\frac{\pi}{6}\right)^{\frac{1}{3}} C_D \left(\frac{4\pi}{3}\right)^{-\frac{\beta}{3}} \left(\frac{\rho_p}{\rho_t}\right)^{\frac{1}{3}} D_p^{1-\beta} g^{-\beta} (v_{imp} \sin\theta_{imp})^{2\beta}, \tag{10}$$

where $C_D = 1.4$, $\beta = 0.17$, $\rho_p = 2.7 \times 10^3$ kg/m$^3$, $\rho_t = 2 \times 10^3$ kg/m$^3$, and $g = 0.0057$ m/s$^2$ for Phobos and $g = 0.003$ m/s$^2$ for Deimos are a dimensionless scaling constant, dimensionless scaling exponent, projectile density, target density, and gravitational acceleration, respectively. The scaling parameters correspond to the values for dry sand (Schmidt and Housen, 1987). The Mars rocks were assumed to be basaltic. The bulk density of the regolith on the moons was estimated by Andert et al. (2010). A typical value of $D_f$ in the Monte Carlo calculations was ~10 m, which is two orders of magnitude larger than $D_p$. The small gravitational acceleration on the surfaces of Phobos and Deimos is the main contributor to this size enhancement. The thickness of the collapsed lens $H_c$ can also be estimated using $D_f$ as $H_c \sim 0.1 D_f \sim 1$ m. The $H_c$ value implies that the typical mixing depth of the Mars rocks is ~1 m, which is an order of magnitude larger than $D_p$. The typical mixing ratio of the Mars rocks to the regolith is ~10 ppm, and ranged from ~10 ppb to ~100 ppm in our Monte Carlo calculations.

Given that the surface area of a collapsed lens is equivalent to that of the host crater, the ratios of the cumulative surface area of the final crater $S_{total} = \sum \pi D_f^2$ to the surface areas of the moon $S_{moon}$ correspond to the access probability to the Mars rock craters $P_{crater}$, when we consider random sampling of the moon regolith somewhere on the surface. In the fiducial case, $P_{crater}$ values on Phobos and Deimos were 3.4% and 0.28%, respectively. Note that the scatter of $P_{crater}$ values in different runs was relatively small (<0.1%). Although the size of each final crater varied from 3 to 12 m, due to the differences in $v_{imp}$ and $\theta_{imp}$ of each impact, the effects of such dispersion on $P_{crater}$ values are canceled out in the summation of all the impacts.

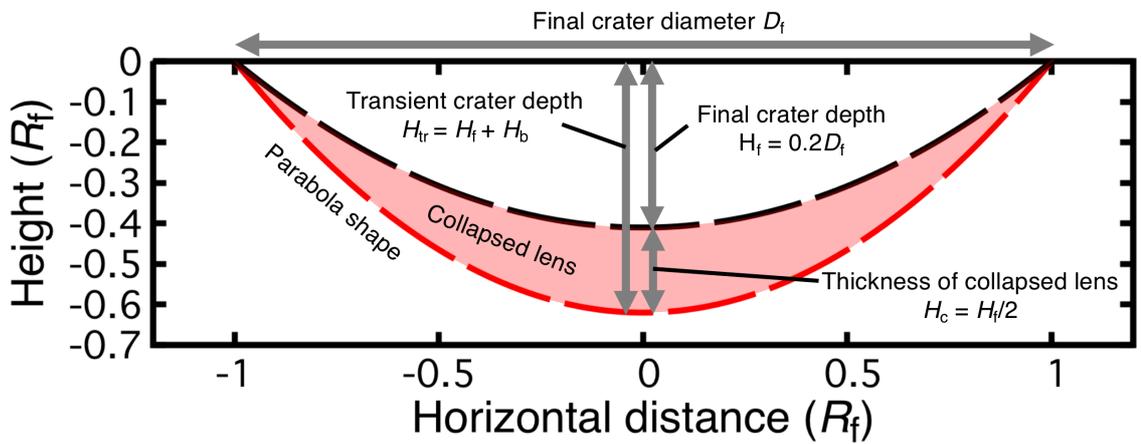

**Figure 8.** Schematic cross-section of Mars rock impact craters considered in this study. The black dashed line is the profile of the final crater that actually formed in the regolith. The red shaded region corresponds to the collapsed lens where the Mars rock fragments mix with the moon regolith particles.

**3.2. Scattered fragments**

In the previous section, we modeled the part (20–30%) of a microbe-bearing Mars rock that mixes into a collapsed lens lying on the floor of a Mars rock crater, based on the experimental results of Daly and Schultz (2016). We now consider the fate of the materials ejected from the Mars rock craters, which represent 70–80% of the mass of the transported Mars rocks. In the case of oblique impacts, projectiles have a large horizontal impact velocity component parallel to the target surface, as compared with the escape velocities $v_{esc}$ of the moons (11 m/s for Phobos and 6.9 m/s for Deimos). In contrast, the escape velocities from the Martian system at the orbits of the moons are the same order of magnitude (3 km/s for Phobos and 2 km/s for Deimos) as the impact velocity of the Mars rocks (Fig. 7a). Therefore, the fragmented projectiles are expected to escape the local gravitational field of each moon and disperse into orbits around Mars. The ejecta from the Mars rock craters are also injected into dust clouds produced at a similar semi-major axis for each moon. Hereafter, the dispersed dust cloud around Mars is referred to as the "torus" in this study, although the dust cloud may have a much wider structure than that recalled by the word "torus" (e.g., Ramsley and Head, 2013). The mass of fragmented projectile $M_{dis,p}$ and escaped mass of regolith particles $M_{dis,t}$ (e.g., Housen et al., 1983) are given by

$$M_{dis,p} = (1 - \psi) M_p, \tag{11}$$

and

$$M_{dis,t} = C_V \rho_t R_{tr}^3 \left(\frac{v_{esc}}{\sqrt{gR_{tr}}}\right)^{-3\mu} \tag{12}$$

where $C_v = 0.32$, $R_{tr} = 0.5 D_{tr}$, and $\mu = 0.4$ are a scaling constant, transient crater radius, and velocity-scaling exponent, respectively, and $v_{esc}$ is the escape velocity from each of the moons (11 m/s for Phobos and 6.9 m/s for Deimos). Equation (12) is derived from the point-source theory (e.g., Holsapple and Schmidt, 1982; Housen et al., 1983). Typical values of $M_{dis,p}$ and $M_{dis,t}$ are ~1 and ~100 kg, respectively, indicating that the mixing ratio of the Mars rock fragments to the regolith from Phobos or Deimos in a dust torus is likely to be ~1%.

The ejected particles then re-accumulate onto Phobos and Deimos (e.g., Ramsley and Head, 2013) because the moons sweep out their own dust torus. It is difficult to accurately estimate the timescale of re-accumulation without detailed numerical simulations considering the ejection process and orbital evolution of the ejected materials. Nevertheless, the timescale may be much longer than one orbital period of Phobos and Deimos (~10 and ~30 h, respectively). In contrast, most of the rocks ejected from the Martian surface hit Phobos and Deimos within the orbital periods of Phobos and Deimos. The Mars rocks that do not reach Phobos and Deimos accrete onto Mars within just one orbital period, because they were launched from the Martian surface. Otherwise, these rocks would escape from the Martian gravitational field if they had very high ejection velocities (>5 km/s). Thus, we assumed that the dispersed materials re-accumulated onto the uppermost surface of Phobos and Deimos after cessation of the Mars rock bombardment.

The dispersion and re-accumulation of dust particles forms a thin global layer on each of the moons. The thicknesses of the thin layers on each of the moons was estimated to be 30 μm for Phobos and 1 μm for Deimos, by using the total masses of the dust tori (~100$M_{p,total,ave}$) and surface areas of each of the moons. In the above calculations, we assumed a homogeneous dispersion of infinitesimal particles onto the surfaces of Phobos and Deimos. This thickness estimate, however, is expected to be unlikely because such small particles (~1 μm) are removed from the Martian system prior to re-accumulation, within several hours, due to radiation pressure (e.g., Hyodo et al., 2018; Ramsley and Head, 2013). Given that the actual sizes of the particles in the dust torus are highly uncertain, we assumed that the thickness of the thin microbial layer is ~0.1 mm. This assumption does not change the conclusions of this study, unless the thickness is >0.3 mm, because the time constant $TC$ for radiation-induced sterilization is nearly constant (71 yr) for a layer with a thickness of <0.3 mm (Patel et al., 2017; Section 2.3).

### 3.3. Effects of impacts by natural meteoroids

In the previous section, we showed that the escaped Mars rocks from the host craters are expected to form a thin global layer with a thickness of <0.1 mm. Figure 4 clearly shows that the thin microbial layer on the uppermost surface experiences rapid sterilization within ~2 kyr. Although 70–80% of the Mars rocks ejected from the Zunil crater are likely to be mixed into this layer, this major fraction will now be completely sterilized. However, the microbes in this thin layer could have survived, if the microbes were covered by a thick ejecta deposit at some time in the past. Such a "radiation shield" could have been produced by "ordinary background impacts", which are collisions of natural meteoroids with the moons. In this section, we discuss the background impact flux onto each of the moons to assess the role of radiation shielding by ejecta blanketing.

The impact flux of natural meteoroids onto Phobos and Deimos can be estimated from the impact flux onto Mars. The SFD of craters on the surface of Mars is well established (e.g., Hartmann, 2005). The impactor SFD can be estimated from the crater SFD by using the relationship between the impact/target conditions and final crater diameter. The transient crater diameter $D_{tr}$ on Mars can also be estimated using the π-group scaling law (Eq. (10)). Note that the density of the Martian crust is assumed to be $2.7 \times 10^3$ kg/m³, which is the same value as the impactor used in the calculation. To convert from $D_{tr}$ to $D_f$ on Mars, we used two empirical laws, including Eq. (9). Another law is as follows:

$$D_f = 1.2 \, D_c^{-0.13} \, D_{tr}^{1.13}, \qquad (13)$$

where $D_c$ is the transition diameter from a simple to complex crater (e.g., McKinnon et al., 1991). We used the larger $D_f$ calculated from the two equations because we needed to examine $D_f$ against a wide range of $D_p$. Equation (13) could not be applied to small craters, although it is suitable for estimating the size of complex craters (McKinnon et al., 1991). The impactor SFDs for Phobos and Deimos were approximated by the products of the impactor SFD on Mars and the ratios of the geometric cross-sections of the moons to that of Mars (e.g., Ramsley and Head, 2013). Figure 9 shows the impactor SFDs pertaining to Mars, Phobos, and Deimos at a surface age of 0.1 Ma. To convert the crater SFD on Mars to the impactor SFD, an average impact velocity and angle from Mars were used (14 km/s; Ito and Malhotra, 2006; 45° from the tangent plane; Shoemaker, 1962). The SFD can be approximated by a power law as follows:

$$N(>D_p) = D_{pmax}^{2.5} D_p^{-2.5}, \qquad (14)$$

where $N(>D_p)$ and $D_{pmax}$ are the cumulative number of impactors with a diameter larger than $D_p$ and the maximum diameter of the impactors, respectively. The maximum diameter $D_{pmax}$ is defined as the diameter at $N(>D_p) = 1$. The $D_{pmax}$ values for Phobos and

Deimos in the past 0.1 Myr were estimated to be 2.6 and 1.6 m, respectively. Although there is the possibility that larger impactors have struck the moons, such events are statistically rare. Given that the total number of impacts of bodies smaller than 0.1 m is unknown, we decided to extrapolate Eq. (14) to smaller diameters. This extrapolation is regarded reasonable because the exponent of –2.5 is close to the steady-state SFD in a self-similar collision cascade (Tanaka et al., 1996).

The $v_{imp}$ distribution of natural meteoroids can be approximated by a Rayleigh distribution (e.g., Parkos et al., 2018; Zahnle et al., 2003). The most likely impact velocities onto Phobos and Deimos were estimated to be 13.4 and 13.2 km/s, respectively, based on the average impact velocity onto Mars (14 km/s), escape velocity from Mars (5 km/s), and escape velocities of the Martian system from the moon orbits (Hirata, 2017; Schmedemann et al., 2014). We used the well-known $\sin(2\theta_{imp})$ function for the impact angle distribution (Shoemaker, 1962) for both Phobos and Deimos.

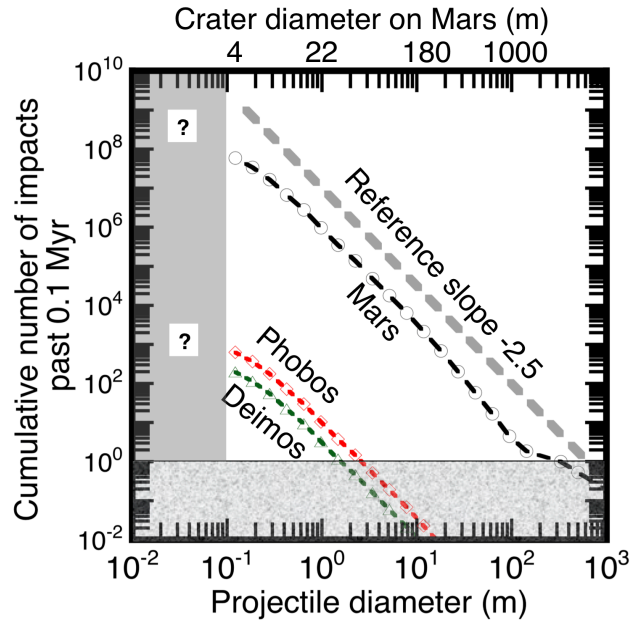

**Figure 9.** Cumulative size–frequency impactor distributions (SFDs) pertaining to Mars, Phobos, and Deimos. The gray dashed line is a reference power law function with an exponent of –2.5. The corresponding crater diameters on Mars are shown on the upper x-axis. Given that the cumulative number of impactors $N$ smaller than 0.1 m is unknown, this region is highlighted in grey and with question marks. The diameters at $N = 1$ correspond to the maximum size of impactors during the past 0.1 Ma.

To investigate the significance of the ejecta deposit covering the thin microbe-bearing layer, knowledge of the rate of decrease of the ejecta deposit thickness with increasing distance along the surface from the host crater is necessary. If the thin microbial layer was covered by an ejecta deposit with a thickness of >3 mm within 2 kyr of the Zunil-forming event, then the microbes might survive for 0.1 Myr (i.e., until now) (Fig. 4). Thus, in the remainder of this section we estimate the total coverage of an ejecta deposit layer with a thickness of >3 mm produced by natural meteoroid impacts. We also used the π-group scaling laws in the gravity-dominated regime in the estimation.

The mass of ejecta in a velocity bin ($v_{ej}$, $v_{ej}+dv_{ej}$) is given as follows (e.g., Housen et al., 1983):

$$m_{ej}(v_{ej})dv_{ej} = 3\mu C_V \rho_t R_{tr}^3 (gR_{tr})^{\frac{3\mu}{2}} v_{ej}^{-3\mu-1} dv_{ej}, \tag{15}$$

where $v_{ej}$ and $dv_{ej}$ are the ejection velocity and a small increment of $v_{ej}$, respectively. The values of $\mu$ and $C_v$ are the same as used in Eq. (12). The ballistic range of the ejecta in bin $R_b$, including the curvature effect, is as follows (e.g., Melosh, 1989):

$$R_b = 2R_{moon}\tan^{-1}\left[\frac{\left(\frac{v_{ej}^2}{gR_{moon}}\right)\sin\theta_{ej}\cos\theta_{ej}}{1-\left(\frac{v_{ej}^2}{gR_{moon}}\right)\cos^2\theta_{ej}}\right], \tag{16}$$

where $R_{moon}$ and $\theta_{ej}$ are the radius of Phobos or Deimos and the launch angle of the ejecta, respectively. We used $\theta_{ej} = 45°$ measured from the target surface in this calculation (e.g., Melosh, 1989). The surface area of the ejecta landing site for the velocity bin $S_{ej}$ is given by

$$S_{ej} = 2\pi R_{moon}\sin\lambda d\lambda, \tag{17}$$

where

$$\lambda = \frac{R_b}{R_{moon}} \tag{18}$$

is the arc angle between the impact point and ballistic range measured from the center of each of the moons. Consequently, the thickness of the ejecta deposit launched in the velocity bin $L_{ej}$ can be obtained as

$$L_{ej} = \frac{m_{ej} dv_{ej}}{S_{ej} \rho_t}. \tag{19}$$

Figure 10 shows examples of the ejecta thickness on Phobos (g = 0.0057 m/s$^2$ and $R_{moon}$ = 11 km) as a function of distance from the impact point. We show the results for four different $D_p$ values. The impact velocity and angle were set to the typical values (13.4 km/s and 45°). We set the ejecta thickness to be zero at a distance shorter than the final crater radius. The ejecta thickness decreases with increasing distance following a power law with an exponent of −2.6. The exponent is consistent with the results of explosion experiments and numerical calculations (e.g., Melosh, 1989). We found that projectiles larger than $D_p$ > 1 mm could produce a thick ejecta deposit layer with a thickness of >3 mm. Thus, the minimum diameter of the natural meteoroids $D_{pmin}$ was set to 1 mm. Although our model produces an antipodal focusing of the ejecta, the effects can be neglected because the thickness at the antipodal point is much less than 3 mm, even when we considered a relatively large impactor with $D_p$ = 1 m. Such large impacts with $D_p$ > 1 m statistically occur every 10 kyr on Phobos (Fig. 9). The total number of impactors $N_{total,BG}$ within 2 kyr can be obtained from the impactor SFD and values of $D_{pmax}$ and $D_{pmin}$ as follows:

$$N_{total,BG} = \left(\frac{2 \text{ kyr}}{0.1 \text{ Myr}}\right) D_{pmax}^{2.5} D_{pmin}^{2.5}, \tag{20}$$

Given that the impact rate on the Martian system in the past 3 Gyr is roughly constant (e.g., Neukum et al., 2001; Schmedemann et al., 2014), we simply multiply the cumulative number of impactors within 0.1 Myr by the ratio $t_{req}$ for the uppermost surface to the time after the Zunil-forming event to the formation time of the Zunil crater. The total numbers of natural meteoroids within 2 kyr of the Zunil-forming impact $N_{total,BG}$ obtained for Phobos and Deimos are 6.7 × 10$^6$ and 2.0 × 10$^6$, respectively.

We used the same Monte Carlo model employed in the calculation of the Mars rock bombardment (Section 3.1) to estimate the total coverage of the thick ejecta deposit layer by combining the impactor SFD, $N_{total,BG}$, and distributions of $v_{imp}$ and $\theta_{imp}$. The surface area of the thick ejecta deposit layer $S_{shield}$ is given by

$$S_{shield} = \pi(R_{b,3mm}^2 - R_f^2), \qquad (21)$$

where $R_{b,3\,mm}$ and $R_f$ are the distance from the impact point covered by the thick ejecta deposit layer, along with the surface and final crater radii, respectively. The ratio of the cumulative surface area of the thick ejecta layer $S_{shield}$ to $S_{moon}$ provides the access probability to the shielded microbial layer $P_{layer}$, where $S_{moon}$ is the surface area of each moon ($1.5 \times 10^9$ m$^2$ for Phobos and $5.0 \times 10^8$ m$^2$ for Deimos). We statistically obtained $P_{layer}$ values for Phobos and Deimos of 0.11% and 0.097%, respectively, from the Monte Carlo runs. This means that 99.9% of the microbes in the thin global layer would have been sterilized within ~2 kyr of the Zunil-forming impact event. It should be mentioned here that the globally-averaged thickness of the ejecta deposit formed with in 2 kyr after the Zunil formation by the background impact flux is much thinner than 3 mm. Thomas et al. (2000) reported that the large ejecta blocks up to ~85 m from the Stickney crater have experienced only a few meters of burial or erosion on Phobos. The age of Stickney crater has been estimated to be 100–500 Myr ago (Ramsley and Head, 2017). The maximum deposition rate of the ejecta from these results is constrained to be <60 μm per 2 kyr. Our Monte Carlo model described in this section could also estimate the globally-averaged thickness of the ejecta deposit $L_{ave}$ as $L_{ave} = [\Sigma m_{ej}(<v_{esc})]/ \rho_t S_{moon}$, yielding that $L_{ave}$ = (33±16) μm per 2 kyr on Phobos. We conducted 100 Monte Carlo runs and obtained the uncertainty as 1σ of the results. Thus, our result does not violate the observational constraints about the burial of the ejecta blocks on Phobos.

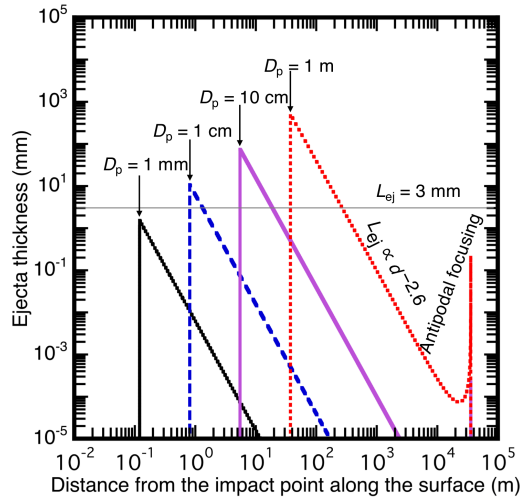

**Figure 10.** Ejecta thickness as a function of distance from the impact point along the surface of Phobos. We used $D_p = 0.001, 0.01, 0.1$, and $1$ m in these calculations. The grey horizontal line corresponds to an ejecta thickness of 3 mm.

## 4. Statistical analysis of microbial contamination probability

In the previous section, we described the fate of the Mars rocks after transportation based on impact physics. The Mars rocks (~0.1 m in diameter) are expected to be disrupted into fine fragments due to the impacts on the moon surface. Some 20–30% of the Mars rock fragments are entrained into "collapsed lenses" on the Mars rock crater floors. The microbes in the retained Mars rock fragments would be well mixed into the collapsed lenses. Given that the thickness of the collapsed lens was estimated to be ~1 m, the microbes retained in the Mars rock craters could have survived radiation sterilization through to the present. The material ejected from the Mars rock craters escaped from each of the moons and re-accumulated on their uppermost surfaces. This forms a thin global layer with a thickness of <0.1 mm. Approximately 0.1% of the area of the thin global layer is shielded from radiation by a thick ejecta deposit layer with a thickness of >3 mm. Hereafter, this covered portion is referred to as "the covered microbial thin layer". In section 4.1, we discuss the microbe concentration of a collapsed lens in a Mars rock crater and the covered microbial thin layer at the present-day. We then assess the sampling probability of the microbes as functions of sampling depth, area, and mass in Sections 4.2 and 4.3. Here, we only use the average value of the transported mass of Mars rocks to the moons in order to clarify the procedure used to calculate the sampling probability. The results of these full statistical analyses are presented after taking into consideration the uncertainties in the input parameters in Section 4.4.

### 4.1. Temporal variations in the number of surviving microbes on the moons

Firstly, we investigated the survival rate $\xi$ (= $N/N_0$) of the microbes after each impact event by using Eq. (3) and the fiducial value for the parameters. The current model of impact sterilization cannot account for a $\theta_{imp}$ dependence on $\xi$. However, it should be noted that the SterLim study (Summers, 2017) performed impact experiments at two different impact angles (40° and 90°) measured from the target surface, and no systematic differences in $\xi$ were observed between the two impact angles. Thus, the average survival fraction $\xi_{ave}$ was simply obtained by a convolution of the $v_{imp}$ distribution with the survival rate $\xi$ given by Eq. (3). We found that $\xi_{ave}$ of the transported microbes for the Zunil crater-forming impact could be much lower than the values assumed in a previous study (Patel et al., 2018). In the cases of Phobos and Deimos, $\xi_{ave}$ values were $2.9 \times 10^{-5}$ and $5.6 \times 10^{-4}$, respectively.

Secondly, we considered the Mars rock craters. The microbe density of the collapsed lens in a Mars rock crater $n_{crater0}$ immediately after crater formation is given by

$$n_{crater0} = \frac{\xi n_{MR} \psi M_p}{(M_c + M_p)}, \qquad (22)$$

where $M_p = 1.4$ kg is the projectile mass (diameter = 0.1 m) and $n_{MR}$ is the microbe density of the Mars rocks immediately prior to arrival at the moons. The fiducial value of $n_{MR}$ is $10^7$ cells/kg, as mentioned in Section 2.4. Due to both impact sterilization and the dilution of microbe-bearing Mars rocks in the collapsed lens, $n_{crater0}$ becomes much less than $n_{MR}$. The average values of $n_{crater0}$ after the bombardment on Phobos and Deimos were $3.1 \times 10^{-3}$ and $4.8 \times 10^{-2}$ cells/kg, respectively. As shown in Fig. 6, the average impact velocity onto Deimos is lower than that onto Phobos, resulting in a higher average $n_{crater0}$ on Deimos than on Phobos. Figure 11 shows the cumulative frequency distributions of $n_{crater0}$ for Phobos and Deimos obtained by the Monte Carlo model. The distributions suggest that for Phobos and Deimos more than 90% and 50%, respectively, of the collapsed lens materials are sterilized down to $10^{-5}$ cells/kg, which corresponds to the REQ-10 criterion for 100 g sampling immediately after crater formation. In contrast, there are Mars rock craters with $n_{crater0} > 1$ cells/kg on Deimos (~2%). Since the total number of Mars rocks for Deimos is $2.7 \times 10^4$ in the fiducial case, the number of craters with >1 cells/kg is estimated to be ~$5 \times 10^2$.

Thirdly, we consider the covered microbial thin layer. The microbial column density immediately after the formation of the thin global layer $\sigma_{thin0}$ can be estimated as follows:

$$\sigma_{thin0} = \frac{\xi_{ave} n_{MR}(1-\psi_{ave}) M_{p,total,ave}}{S_{moon}}. \tag{23}$$

Our model yields $\sigma_{thin0}$ values on Phobos and Deimos of $2.9 \times 10^{-5}$ and $3.3 \times 10^{-5}$ cells/cm$^2$, respectively, in the fiducial case. The shielding by the ejecta deposit layer due to natural meteoroid impacts is now considered. In 2 kyr, which is the survival period from radiation-induced extinction of microbes in the thin global layer on the uppermost surface, the microbial column density $\sigma_{thin}(t)$ rapidly decreases with time as given by

$$\sigma_{thin}(t) = \sigma_{thin0} \exp\left(-\frac{t}{TC}\right) \tag{24}$$

where $t = i\Delta t$ is the time at the $i$-th impact and $\Delta t = 2 \times 10^3$ yr/$N_{total,BG}$. Here, we used $TC$ = 71 yr, which corresponds to the value for a thin layer with a thickness of <0.3 mm (Patel et al., 2018). The number of microbes protected by the ejecta deposit layer $N_{layer}$ at the $i$-th impact can be expressed as

$$N_{layer} = \sigma_{thin}(t) S_{shield}. \tag{25}$$

The ratio $\Sigma N_{layer}/\Sigma S_{shield}$ yields the average microbial column density of the covered microbial thin layer $\sigma_{thin,ave}$ at the present on Phobos and Deimos of $1.0 \times 10^{-6}$ and $1.1 \times 10^{-6}$ cells/cm$^2$, respectively. By comparing with $\sigma_{thin0}$, we found that >95% of the microbes in the thin microbial layer are sterilized before they are covered by the ejecta blanket.

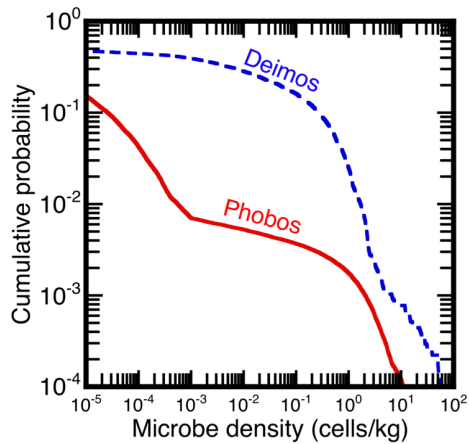

**Figure 11.** Cumulative probability of microbe concentrations in collapsed lenses in Mars rock craters. Results for Phobos (red solid line) and Deimos (blue dashed line) are shown.

We examined the total number of surviving microbes $N_{surv}$ on each of the moons. The total number of surviving microbes in Mars rock craters $N_{surv,crater}$ is approximated as follows:

$$N_{surv,crater} = \eta(0.1 \text{ Myr}, 1 \text{ m}) \xi_{ave} n_{MR} \psi_{ave} M_{p,total}. \quad (26)$$

The total number of microbes protected by the thick ejecta deposit layer $N_{surv,layer}$ can be estimated as follows:

$$N_{surv,layer} = \xi_{ave}(1 - \psi_{ave}) P_{layer}\left(\frac{\sigma_{thin}}{\sigma_{thin0}}\right) n_{MR} M_{p,total}. \quad (27)$$

The total microbe numbers $N_{surv,crater}$ on Phobos and Deimos were estimated to be $3.8 \times 10^7$ and $6.3 \times 10^7$ cells, respectively. Although 70–80% of the transported mass is partitioned into the thin global layer, the contribution of $N_{surv,layer}$ to $N_{surv}$ is much smaller than $N_{surv,crater}$ because the $N_{surv,layer}$ values on Phobos and Deimos were only $1.7 \times 10^4$ and $6.9 \times 10^2$ cells, respectively. This reduction in $N_{surv,layer}$ clearly shows the significance of the radiation-induced sterilization at the uppermost surface. Consequently, the bulk survival rate $N_{surv}/N_{total}$ on Phobos and Deimos are $1.9 \times 10^{-6}$ and $1.7 \times 10^{-4}$, respectively, implying that design of microbe-return missions from either of the Martian moons is quite difficult compared to the case of Mars. A comparison between $N_{surv}$ on Phobos and Deimos leads to an interesting conclusion, in that microbial contamination on Deimos is somewhat greater than that on Phobos, even though Deimos is located much more further from Mars than Phobos.

Figure 12 shows an example of temporal variations in the surviving microbe numbers in the case of Phobos. Since the duration of Mars rock bombardment is quite short, as discussed in Section 3.2, the surviving microbe number suddenly decreases at $\sim 10^{-3}$ yr after the Zunil-forming impact event, which is $\sim 10$ h after the impact. Most of the microbes in the dust torus then re-accumulate onto the uppermost surface within several or several tens of orbital periods. At this time, the surviving microbes are concentrated in the thin global layer. However, radiation-induced sterilization rapidly proceeds to reduce the microbes in the thin layer within $\sim 1$ kyr. The transition from a "global contamination mode" to a "patchy contamination mode" occurs at ca. $10^2$–$10^3$ yr after the Zunil-forming impact. After the transition, the total surviving microbe numbers in the Mars rock craters are three to four orders of magnitude higher than that in the covered microbial thin layer.

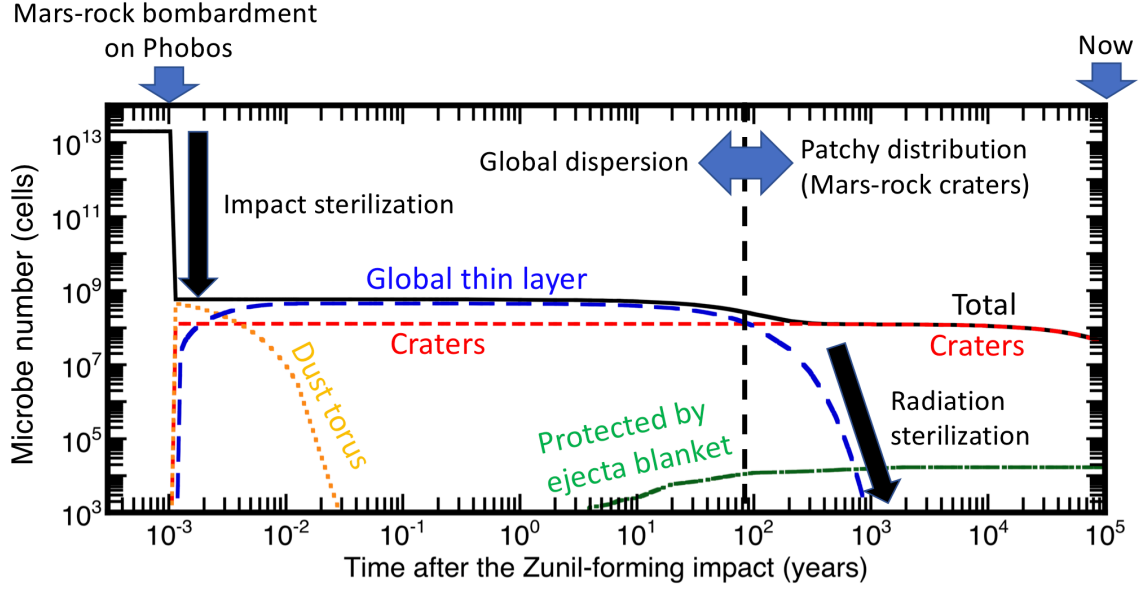

**Figure 12.** Temporal variations in the number of microbes on Phobos. The total number of transported microbes is $2 \times 10^{13}$ cells in this calculation. The total number of surviving microbes at the present is $4 \times 10^7$ cells, which is 2 ppm of the initial value.

### 4.2. Microbial contamination probability from collected samples

In this section, we discuss the sampling probability $P_s$ of the microbes in the case of random sampling from somewhere on each of the moons. According to REQ-10, the boundary between the two categories, which are *Restricted* and *Unrestricted Earth Return* missions, is $P_s = 10^{-6}$, as mentioned in Section 1. The sampling probabilities of the microbes from the Mars rock craters and covered microbial thin layer $P_{s,\text{crater}}$ and $P_{s,\text{layer}}$ are respectively given by

$$P_{s,\text{crater}} = P_{\text{crater}} \eta(t, H) n_{\text{crater0}} M_s, \qquad (28)$$

and

$$P_{s,\text{layer}} = P_{\text{layer}} \sigma_{\text{thin,ave}} S_s, \qquad (29)$$

where $M_s = \rho_t S_s L_s$ is the sample mass, $S_S$ is the sample area, $L_S$ is the sample depth, and $\sigma_{\text{thin,ave}}$ is the average value of the areal microbe density in the covered microbial thin layers. The sum of $P_{s,\text{crater}}$ and $P_{s,\text{layer}}$ is $P_s$. If we choose a sampling mass, the relationship between $S_S$ and $L_S$ is determined uniquely. Figure 13 shows $P_s$ on Phobos and Deimos as functions of sampling area, depth, and mass. Five different sampling masses were used in the calculation. In the fiducial case, we could collect regolith down to ~5 cm for Phobos and ~3 cm for Deimos in the category of the *Unrestricted Earth return* mission, if $M_s$ = 100 g. As discussed in Section 4.4, the above estimates are highly conservative.

If we could avoid the Mars rock craters during the operational phase of sample collection, $P_s$ becomes $P_s = P_{s,\text{layer}}$, as displayed on Fig. 13 as straight lines. In this case, $P_s$ monotonically decreases with increasing sampling depth, because the microbes in the thin layer are diluted by the regolith particles from non-contaminated deeper regions. Thus, the limitation related to sampling depth can be removed. By substituting $P_s = P_{s,\text{layer}} = 10^{-6}$, the allowable sampling surface area on each of Phobos and Deimos can be determined to be $9.1 \times 10^2$ and $9.9 \times 10^2$ cm², respectively, which are obviously large enough to allow the design of boring-type sampling instruments.

**4.3 Contribution of unrecognized crater events**

Given that a global catalog of Martian craters with diameters larger than 1 km has already been compiled (e.g., Robbins and Hynek, 2012), there is little possibility that young Zunil-sized craters have been overlooked. However, it is informative to examine the effects of potential recent crater-forming events on the microbe sampling probability. Here, the diameter of such a putative crater was set to 10 km, which is equivalent to the size of the Zunil crater. The crater chronology model on Mars was statistically constructed based on lunar and Martian crater SFDs (e.g., Neukum et al., 2001). The model allows us to examine the expected number of craters with a given diameter at a given time (e.g., Hartmann, 2005). The number is equivalent to the event probability $P_{\text{event}}$ at a given surface age. We examine the effects of a possible unrecognized Zunil-sized crater on the microbe sampling probability from Phobos by considering the event probability. In a statistical sense, Zunil-sized craters are expected to be formed every ~1 Myr (e.g., Hartmann, 2005). Thus, $P_{\text{event}}$ is only 0.1 if the Zunil crater was actually produced at 0.1 Ma. Although we used 0.1 Ma as the formation age of the Zunil crater to perform a conservative risk assessment, the estimated age of the Zunil crater is 1.0 to 0.1 Ma (e.g.,

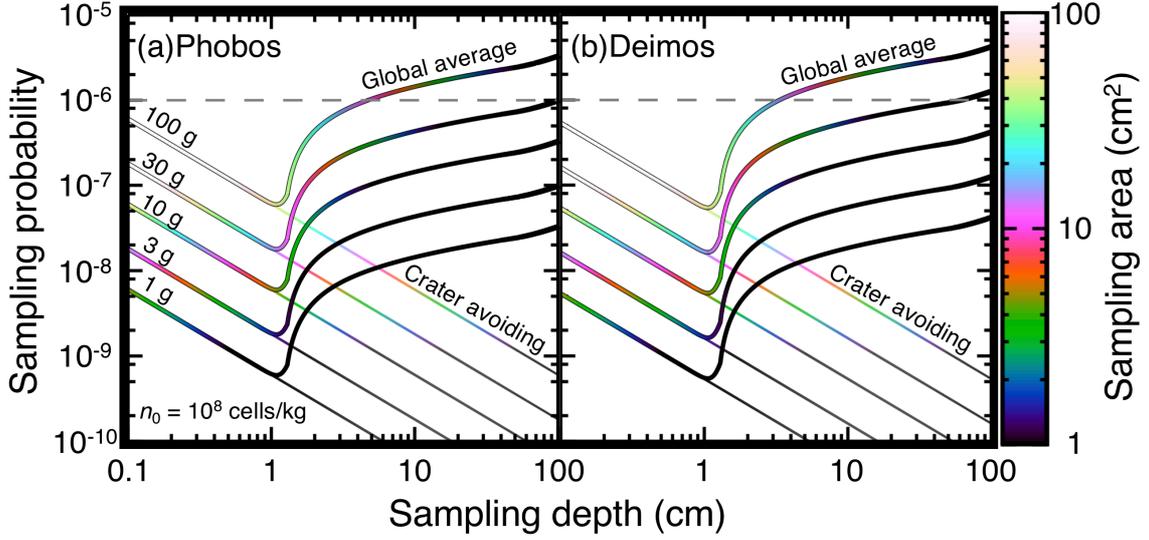

**Figure 13.** Sampling probabilities of microbes from the surfaces of (a) Phobos and (b) Deimos as functions of sampling depth, area, and mass. The selected sampling masses are annotated beside each line. The REQ-10 criterion is shown as a grey dashed horizontal line. The curves are the globally averaged probabilities including the Mars rock craters. The straight lines on the log–log plots are the sampling probabilities excluding the Mars rock craters.

Hartmann, 2010). Since the impact flux onto Mars in the past 3 Gyr is nearly constant (e.g., Neukum et al., 2001), it is straightforward to estimate $P_{event}$ for unrecognized Zunil-sized craters at any surface age as follows:

$$P_{event} = \left(\frac{t}{1~\mathrm{Myr}}\right), \qquad (30)$$

where $t$ is the formation time before the present of the unrecognized crater.

We used the $v_{imp}$ and $\theta_{imp}$ distributions for the Mars rocks from a putative unrecognized crater obtained by orbital calculations (Hyodo et al., in preparation). Since the location of the crater is unknown, the location on Mars of the putative crater was randomly varied in the 10,000 trajectory analyses. The details are presented in Hyodo et al. (in preparation). The average transported mass $M_{p,total,ave}$ of Mars rocks from Zunil-sized craters to Phobos was estimated to be $8.3 \times 10^5$ kg in the random case. By using the $v_{imp}$ and $\theta_{imp}$ distributions pertaining to Phobos, the average survival rate for the impact-induced sterilization $\xi_{ave}$ and average fraction of projectile retention $\psi_{ave}$ were estimated to be $6.7 \times 10^{-4}$ and 0.29, respectively. The sampling probability of microbes transported from an unrecognized crater can be obtained as follows:

$$P_s \sim \begin{cases} P_{event}\sigma_{thin}(t)S_s & (\text{if } t < 2 \text{ kyr}) \quad (31\text{-}1) \\ P_{event}P_{s,crater} + P_{s,layer} & (\text{if } t > 2 \text{ kyr}) \quad (31\text{-}2) \end{cases}$$

If $t < 2$ kyr, the moons are classified as being in the global contamination mode defined in Section 4.1, and the surviving microbes are concentrated in the uppermost, thin global layer, resulting in a depth-independent $P_s$. If we consider the case of Phobos with $M_{p,total,ave} = 8.3 \times 10^5$ kg, $t = 100$ yr, and $S_s = 3$ cm², $P_s$ is estimated to be $2 \times 10^{-8}$. It is notable that $P_s$ becomes lower than $10^{-6}$ even if $P_{event} > 10^{-6}$ due to the impact-induced sterilization and dilution of the microbes across the entire surface of the moons. Thus, the contamination risk by unrecognized extremely young craters can be neglected. In Eq. (31-2), $P_{s,layer}$ is not relevant to the formation time $t$ of the crater, because it is determined only by the background impacts within 2 kyr after the Zunil-forming impact event. Figure 14 is the same as Fig. 13, except that it shows the calculated results for an unrecognized Zunil-sized crater at six different formation ages from 3 ka to 1 Ma, $M_s = 60$ and 100 g are only displayed, and only results for Phobos are provided. Given the survival rate from the radiation $\eta(t, H)$ and $P_{event}(t)$ are competitive in the sampling probability calculation, the sampling probability exhibits a complex behavior with changing formation time $t$. Nevertheless, we can rule out a risk of microbial contamination from an unrecognized crater down to ~10 cm below the surface of Phobos, if the sampling mass is limited to <60 g.

**4.4. Propagation of uncertainties to the microbial contamination probability**

In this section, we discuss the potential uncertainties in the microbial contamination probability. Here, we only consider the case of microbial contamination of Phobos by the Zunil-forming impact event. We have already shown that the degree of contamination on Deimos is similar to that on Phobos (Section 4.1) and that the effects of unrecognized craters are not significant (Section 4.3). Firstly, we address the effects of the PDF on the initial microbe density, transported mass, and degree of impact sterilization in Section 4.4.1. Secondly, we discuss the effects of the sizes of the Martian rocks. Although all the microbe-bearing Mars rocks are assumed to be spheres that are 0.1 m in diameter based on Artemieva and Ivanov (2004) and our own aerodynamic analyses (Paper 1), we consider how the size of the Mars rocks affects $P_s$ in Section 4.4.2. Finally, we assess how a longer time constant $TC$ affects $P_s$ in Section 4.4.3, because we cannot completely rule out the possibility that there is an unknown microbe more resistant to radiation than the MS-2 coliphage, which has the best radiation resistance of known microbes on Earth.

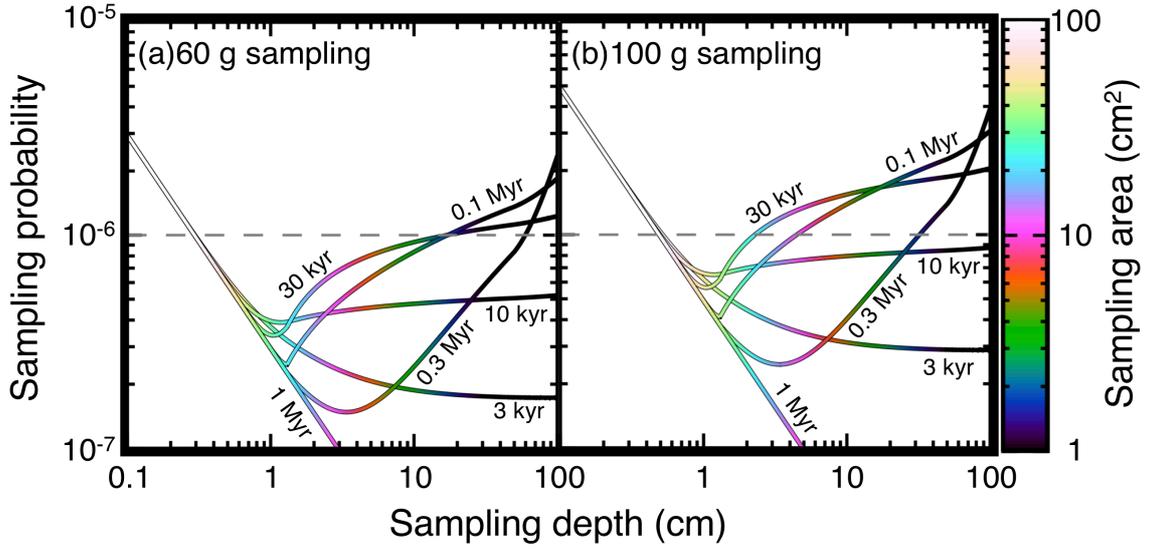

**Figure 14.** Same as Fig. 13 except that the calculated results are for an undiscovered crater. The calculated result in the case of $t > 2$ kyr (Eq. (31-2)) is shown here. (a) $M_s = 60$ g and (b) $M_s = 100$ g. The formation ages are shown beside the lines.

**4.4.1. Full statistical analysis**

In Section 4.2, we showed that the contribution of the microbial thin layer to the sampling probability $P_s$ can be neglected, if we chose a sampling depth of >1 cm. In this case, the full description of $P_s$ is as follows:

$$P_s = n_{\text{Mars}} \alpha \beta_{\text{ave}} \xi_{\text{ave}} \psi_{\text{ave}} \eta(t = 0.1 \text{ Myr}, H = L_s) P_{\text{crater}}(M_{p,\text{total}}) M_s, \qquad (32)$$

where $\alpha$ and $\beta_{\text{ave}}$ are the survival rate during launch and average of the mixing ratio of Mars rocks to collapsed lenses, respectively. Consequently, $P_s$ is expressed simply as the product of the eight parameters, showing that $P_s$ is linearly proportional to these parameters. The only non-linear term is the survival rate for radiation $\eta$ in terms of the change in sampling depth. From the eight parameters, $\alpha$, $\beta_{ave}$, and $\psi_{ave}$ are largely unchanged from the fiducial case used to obtain the $P_s$ curves shown in Fig. 13. We set $\alpha$ = 0.1 as a conservative estimate, although $\alpha$ is likely to be much smaller than 0.1, as discussed in Paper I. The values for $\beta_{ave}$ and $\psi_{ave}$ are tightly constrained from our knowledge of impact physics (Section 3.1), with the $v_{imp}$ and $\theta_{imp}$ distributions obtained from orbital calculations (Hyodo et al., in preparation). Given that $L_s$ was assumed, we can obtain $\eta$ from Fig. 5. The sampling mass $M_s$ can be arbitrarily selected by changing the sampling area $S_S$. The other variables, $n_{Mars}$, $M_{p,total}$, $\xi_{ave}$, and $P_{crater}$, can be changed by more than one order of magnitude from the fiducial case. In this section, we obtained the PDF of $P_s$ and the expected value of $P_s$ for $n_{Mars}$, $M_{p,total}$, $\xi_{ave}$, and $P_{crater}$ by using a Monte Carlo technique.

We now consider the distribution of $M_{p,total}$ and $P_{crater}$. Although we used the average values of $M_{p,total}$ to calculate $P_s$ shown in Fig. 13, >80% of orbital calculations yielded smaller values of $M_{p,total}$ (Fig. 6). In this full statistical analysis, we extracted a single value of $M_{p,total}$ from this distribution in each Monte Carlo run. The access probability to the Mars rock craters $P_{crater}$ should correlate with $M_{p,total}$ and the size of the Mars rocks. We can obtain $P_{crater}$ by using the stochastic impact model used in Section 3. Figure 15 shows $P_{crater}$ as a function of $M_{p,total}$. Although the effects of the $v_{imp}$ and $\theta_{imp}$ distributions are included in the model, we found that $P_{crater}$ can be approximated well by a simple linear function of $M_{p,total}$, due to the averaging effect of a number of Mars rock impact events. Thus, we can easily calculate $P_{crater}$ from the selected $M_{p,total}$ in each Monte Carlo run.

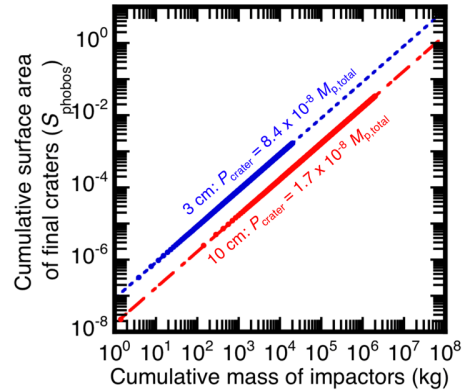

**Figure 15.** Access probability to the Mars rock craters on Phobos as a function of the total mass of transported Mars rocks. Two different sizes (0.03 m in blue; 0.1 m in red) were used as the rock diameters. Note that the actual results obtained from the cratering calculations are shown as filled circles. The dotted lines are the best-fit lines. The formulae for each best-fit line are annotated beside the lines.

We now briefly describe the input distributions of $n_{Mars}$ and $\xi_{ave}$. Although we used a potential microbe density on Mars of $10^8$ cells/kg as the fiducial case throughout this study, the potential microbe density in the Martian environment may be approximated by a Gaussian curve with an average of $10^{7.5}$ cells/kg and standard deviation of $10^{0.5}$ cells/kg, as mentioned in Section 2.4 and Paper 1. Thus, we extracted one value of $n_{Mars}$ from the Gaussian curve in each Monte Carlo run. As discussed in Section 2.2, we constructed a physical model (see Eq. (3)) to constrain the degree of impact sterilization based on the data from the SterLim impact test. We chose a $C$ value from Eqs (2) and (3) for a given Zunil-forming impact event in each Monte Carlo run. We then calculated the average survival rate due to impacts $\xi_{ave}$ on the surface of Phobos in each run by convolving with the $v_{imp}$ distribution obtained from the orbital calculations (Hyodo et al., in preparation).

Figure 16 shows the results of the full statistical analysis, which is the PDF of $P_s$, showing that $P_s$ ranges from $10^{-11}$ to $10^{-5}$ and that the most likely value of $P_s$ is $10^{-8}$, except if the sampling depth $L_s$ is <2 cm. Sample collection from Phobos could satisfy the REQ-10 criterion with 99% confidence even when we consider the case of $M_s$ = 30 g and $L_s$ = 10 cm. Table 1 lists the expected values of $P_s$ as a function of sampling depth $L_s$. Although $P_s$ increases monotonically with $L_s$, $P_s$ becomes <$10^{-6}$ at any $L_s$ value even with 100 g sampling. Consequently, our full statistical analysis strongly suggests that the microbial contamination probability is well below the REQ-10 criterion.

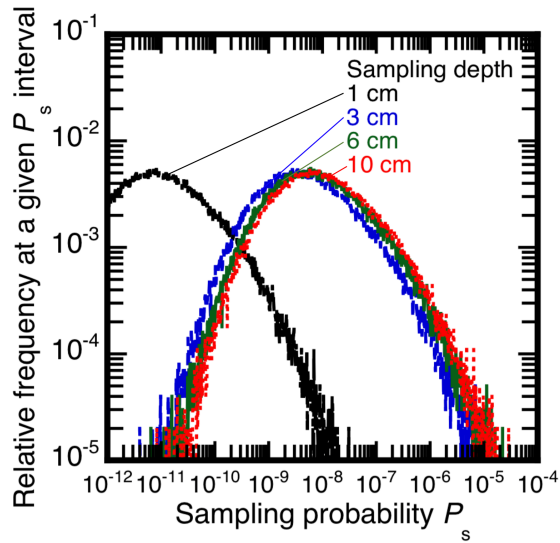

**Figure 16.** Probability density function (PDF) of the microbial contamination probability $P_s$. The sampling mass was set to 30 g. We calculated $P_s$ for four different sampling depths (i.e., 1, 3, 6, and 10 cm). We obtained the PDF via 100,000 Monte Carlo runs.

**Table 1.** Expected values of the microbial contamination probability at a given sampling depth and mass.

| Sampling depth (cm) | 30-g sampling | 100-g sampling |
|---|---|---|
| 1 | 8.5 x $10^{-11}$ | 2.8 x $10^{-10}$ |
| 2 | 2.2 x $10^{-8}$ | 7.5 x $10^{-8}$ |
| 3 | 3.9 x $10^{-8}$ | 1.3 x $10^{-7}$ |
| 6 | 6.3 x $10^{-8}$ | 2.1 x $10^{-7}$ |
| 10 | 7.9 x $10^{-8}$ | 2.6 x $10^{-7}$ |

**4.4.2. Effects of the size of Mars rocks**

In this section, we discuss the sensitivity of expected $P_s$ to the diameter of Mars rocks $D_p$. Although $D_p$ = 0.1 m has been estimated to be the minimum size required for penetrating the Martian atmosphere (Artemieva and Ivanov, 2004; Paper I), these aerodynamic analyses assumed a static atmosphere. It is known that hypervelocity impacts in an atmosphere strongly disturb the flow field of the pre-existing atmosphere (e.g., Quintana et al., 2018; Schultz, 1992; Schultz and Quintana, 2017; Shuvalov, 2009; Sugita and Schultz, 2002). If the disturbed atmosphere moves along with the Mars rocks in a similar direction, then the relative velocities between the ejected Mars rocks and disturbed atmosphere become somewhat smaller with respect to the aerodynamic analyses described in Paper 1, possibly resulting in smaller rocks escaping into space. Since a smaller $D_p$ leads to a higher $P_{crater}$, a smaller $D_p$ also yields a higher $P_s$ (Eq. (32)). As noted in Paper I, Mars rocks having a diameter of <0.03 m are sterilized due to the thermal conduction from the heated atmosphere (several thousand K), even when the relative velocity between the moving rocks and surrounding atmosphere is nearly zero.

Thus, we set the minimum size to be 0.03 m in this sensitivity analysis. In Fig. 15, we also show $P_{crater}$ at $D_p = 0.03$ m as a function of $M_{p,total}$. We confirmed that $P_{crater}$ can also be approximated as a linear function. The comparison between the proportional coefficients shows that $P_{crater}$ at $D_p = 0.03$ m is five times greater than that at $D_p = 0.1$ m, at a given $M_{p,total}$, showing that $P_s$ is $<10^{-6}$ for <10 cm deep and 30 g sampling (Table 1).

### 4.4.3. Effects of the time constant on radiation-induced sterilization

We now discuss a putative case where the potential microbes on Mars are more radiation resistant than the MS-2 coliphage. The change in $P_s$ for such microbes can be roughly estimated by using longer time constants $TC$ for radiation-induced sterilization. Table 2 shows the survival rate for radiation at a given time and depth $\eta(t, H)$ with different $TC$ values. We consider microbes with three- and six-fold higher $TC$ values at $t = 0.1$, 2, and 3 Myr. The times $t$ correspond to the minimum estimates of the formation of the known large craters on Mars, such as Zunil (10.1 km, $t = 0.1$ Ma), Tooting (29 km, $t = 2$ Ma), and Mojave (58 km, $t = 3$ Ma). At $t = 0.1$ Ma and $H = 10$ cm, $\eta$ is calculated to be in the range 14% to 66%, suggesting that radiation-induced sterilization largely does not proceed after the transported microbes are embedded in the collapsed lenses. Thus, the change in the expected $P_s$ value for the unknown microbes, which are much more resistant against radiation than the MS-2 coliphage, is at most a factor of five even for a six-fold increase in $TC$ values and $L_s = 10$ cm.

We now consider the effects of microbes with a longer $TC$ from Tooting and Mojave craters on $P_s$. The ratio of $\eta$ at $t = 0.1$ Ma to $\eta$ at 2 Ma is >300 regardless of the depth and $TC$. In the same manner, the ratio of $\eta$ at $t = 0.1$ Ma to $\eta$ at 3 Ma is >6,000. Although the average values of $M_{p,total,ave}$ from Mojave and Tooting are calculated to be 200 and 20 times greater than $M_{p,total}$ from Zunil, respectively (Hyodo et al., in preparation), the mass contrast is much smaller than the radiation survival rate. As such, the microbes in the collapsed lenses that were transported from the Mojave and Tooting craters are now largely sterilized and their contribution to $P_s$ can be neglected.

**Table 2** The radiation survival rate integrated over a given depth. Three different *TC* values are used here.

| Sampling depth (cm) | $t = 0.1$ Myr (Zunil) | | | $t = 2$ Myr (Tooting etc.) | | | $t = 3$ Myr (Mojave) | | |
|---|---|---|---|---|---|---|---|---|---|
| | Normal | × 3 | × 6 | Normal | × 3 | × 6 | Normal | × 3 | × 6 |
| 1 | $1.5 \times 10^{-4}$ | $2.0 \times 10^{-2}$ | $8.7 \times 10^{-2}$ | 0 | $2.3 \times 10^{-20}$ | $3.3 \times 10^{-11}$ | 0 | $1.8 \times 10^{-29}$ | $8.4 \times 10^{-16}$ |
| 3 | $6.6 \times 10^{-2}$ | $3.0 \times 10^{-1}$ | $4.7 \times 10^{-1}$ | $2.4 \times 10^{-19}$ | $3.4 \times 10^{-7}$ | $4.2 \times 10^{-4}$ | $2.4 \times 10^{-28}$ | $2.9 \times 10^{-10}$ | $1.2 \times 10^{-5}$ |
| 6 | $1.1 \times 10^{-1}$ | $4.2 \times 10^{-1}$ | $6.0 \times 10^{-1}$ | $2.9 \times 10^{-17}$ | $1.8 \times 10^{-6}$ | $1.1 \times 10^{-3}$ | $3.0 \times 10^{-25}$ | $3.4 \times 10^{-9}$ | $4.5 \times 10^{-5}$ |
| 10 | $1.4 \times 10^{-1}$ | $4.8 \times 10^{-1}$ | $6.6 \times 10^{-1}$ | $7.0 \times 10^{-16}$ | $5.6 \times 10^{-6}$ | $2.0 \times 10^{-3}$ | $3.5 \times 10^{-23}$ | $1.7 \times 10^{-8}$ | $1.0 \times 10^{-4}$ |

## 5. Conclusions

We investigated the surface processes on the moons Phobos and Deimos after the potential transportation and arrival of Martian microbes from a young ray crater (Zunil) on Mars by using the results of Paper 1 (Fujita et al., in submission) as the initial conditions. In this study, we used the orbital parameters of Mars rocks, including impact velocity and angle, and the transported mass to each of the moons, obtained by two companion studies (Genda et al., in preparation; Hyodo et al., in preparation). The processes driven by Mars rock impacts were investigated based on impact physics. The transported Mars rocks produce many craters on the regolith layers of the moons. The Mars rocks are broken into fine fragments due to the stress pulse during impact. Some of the fragments are retained and mixed into the collapsed regolith layer on the floor of the final crater. Some material escapes from the impacted moon and forms a thin global layer. The subsequent radiation-induced sterilization was also taken into account. We identified two types of location with a high microbe density: (1) collapsed regolith layers in Mars rock craters; and (2) thin microbe layers covered by a thick ejecta layer produced by natural meteoroid impacts onto the moons. The sampling probability of the transported microbes from the former location is much greater than that from the latter one. The surviving microbe fractions transported from Mars are presently only ~1 ppm on Phobos and ~100 ppm on Deimos, suggesting that the design of a microbe sample-return mission from the moons is much more difficult than that from Mars. We calculated the sampling probability of the transported microbes from a randomly selected point on the moons, in the context of planetary protection issues. This showed that regolith samples down to ~5 cm for Phobos and ~3 cm for Deimos could be collected in the category of a *Unrestricted Earth-Return* mission, in the fiducial case. The PDF of the microbial contamination probability was also evaluated by considering the uncertainties on the input variables. The most likely microbe sampling probability value is $10^{-8}$. Samples collected from Phobos could satisfy the REQ-10 criterion with 99% confidence, even considering the case of 30 g and 10 cm deep sampling.

**Acknowledgements**

We thank Hiroki Senshu, Koji Wada, Masahiko Sato, Tomohiro Usui, Fumi Yoshida, Hideaki Miyamoto, and the Phobos–Deimos Microbial Contamination Assessment Team for useful discussions. We appreciate two anonymous referees for their constructive comments that helped greatly improve the manuscript, and Francois Raulin for handling the manuscript as an editor. K.K. is supported by JSPS KAKENHI grant numbers JP17H01176, JP17H01175, JP17K18812, JP18HH04464, JP19H00726 and by the Astrobiology Center of the National Institute of Natural Sciences, NINS (AB301018). H.G. and K. K. are supported by JP17H02990. H.G. is also supported by JP17H06457. R.H. is supported by JSPS Grants-in-Aid (JP17J01269, JP18K13600)

**References**

Andert, T.P., Rosenblatt, P., Pätzold, M., Häusler, B., Dehant, V., Tyler, G.L., and Marty, J.C. (2010) Precise mass determination and the nature of Phobos. Geophys Res Lett 37, doi:10.1029/ 2009GL041829.

Artemieva, N. and Ivanov, B. (2004) Launch of meteorites in oblique impacts. Icarus 171, 84–101.

Barney, B. L., Pratt, S. N., and Austin, D. E., 2016. Survivability of bare, individual Bacillus subtilis spores to high-velocity surface impact: Implications for microbial transfer through space. Planetary and Space Science, 125, 20-26.

Barnouin-Jha, O.S., Yamamoto, S., Toriumi, T., Sugita, S., and Matsui, T. (2007) Non-intrusive measurements of crater growth. Icarus 188, 506-521.

Bogard, D.D. and Johnson, P. (1983) Martian gases in an Antarctic meteorite?. Science 221 651-654.

Burchell, M. J., Mann, J. R., Bunch A. W., and Brandão, F. B., 2001. Survivability of Bacteria in hypervelocity impact. Icarus, 154, 545-547.

Burchell, M. J., Galloway, J. A., Bunch, A. W., and Brandão, F. B. 2003. Survivability of bacteria ejected from icy surfaces after hypervelocity impact. Astrobiology, 33, 53-74.

Burchell, M. J., Mann, J. R., and Bunch, A. W., 2004. Survival of bacteria and spores under extreme shock pressures. Mon. Not. R. Astron. Soc., 352, 1273-1278.

Chappaz, L., Melosh, H.J., Vaquero, M., and Howell, K.C. (2013) Transfer of impact ejecta material from the surface of Mars to Phobos and Deimos. Astrobiology 13, 963-980.

Daly, R.T. and Schultz, P.H. (2016) Delivering a projectile component to the vestan regolith. Icarus 264:9-19.

DeCarli, P.S., Goresy, A.El, Xie, Z., Sharp, T.G. (2007) Ejection mechanisms for Martian Meteorites. In: AIP Conference Proceedings, 955, pp. 1371–1374. doi:10.1063/1. 2832979.

Ebert, M., Hecht, L., Deutsch, A., Kenkmann, T., Wirth, R., and Berndt, J. (2014) Geochemical processes between steel projectiles and silica-rich targets in hypervelocity impact experiments. Geochimica et Cosmochimica Acta 133, 257-279.

Elbeshausen, D., Wunnemann, K., and Collins, G.S. (2013) The transition from circular to elliptical impact craters. Journal of Geophysical Research 118, 1-15, doi:10.1002/2013JE004477.

Eugster, O. (1989) History of meteorites from the Moon collected in Antarctica. Science 245, 1197-1202, DOI: 10.1126/science.245.4923.1197.

Fajardo-Cavazos, P., Langenhorst, F., Melosh, H. J., and Nicholson, W.L. (2009) Bacterial spores in granite survive hypervelocity launch by spallation: Implications for lithopansperimia. Astrobiology. 9, 647–657.

Fiers, W., Contreras, R., Duerinck, F., Haegeman, G., Iserentant, D., Merregaert, J., Min Jou, W., Molemans, F., Raeymaekers, Van den Berghe, A., Volckaert, G., and Ysebaert, M. (1976) Complete nucleotide sequence of bacteriophage MS2 RNA: primary and secondary structure of the replicase gene. Nature, 260, 500-507.

Fujita K., Kurosawa, K., Genda, H., Hyodo, R., Matsuyama, S., Yamagishi, A., Mikouchi, T., and Niihara, T. Assessment of microbial contamination probability for sample return from Martian moons 1: The departure of the microbes from Martian surface. Accepted, *Life Sciences in Space Research*.

Fukuzaki, S. Sekine, Y., Genda, H., Sugita, S., Kadono, T., and Matsui, T. (2010) Impact-induced $N_2$ production from ammonium sulfate: Implications from the origin and evolution of $N_2$ in Titan's atmosphere. Icarus 209, 715-722.

Genda, H., Kobayashi, H., and Kokubo, E. (2015). Warm debris disks produced by giant impacts during terrestrial planet formation. Astrophys. J., 810, 136.

Genda, H., Kurosawa, K., Okamoto, T., Hydrocode modeling of the spallation process during hypervelocity oblique impacts. in prep.

Gillon M., Jehin, E., Lederer, S.M., Delrez, L., de Wit, J., Burdanov, A., Van Grootel, V., Burgasser, A.J., Triaud, A.H.M.J., Opitom, C., Demory, B.-O., Sahu, D.K., Gagliuffi, D.B., Magain, P., and Queloz, D. (2016) Temperate Earth-sized planets transiting a nearby ultracool dwarf star. Nature 533, 221-224.

Gillon M., Triaud, A.H.M.J., Demory, B.-O., Jehin, E., Agol, E., Deck, K.M., Lederer, S.M., de Wit, J., Burdanov, A., Ingalls, J.G., Bolmont, E., Leconte, J., Raymond, S.N., Selsis, F., Turbet, M., Barkaoui, K., Burgasser, A., Burleigh, M.R., Carey. S.J., Chaushev, A., Copperwheat, C.M., Delrez, L., Fernandes, C.S., Holdsworth, D.L., Kotze, E.J., Van Grootel, V., Almleaky, Y., Benkhaldoun, Z., Magain, P., and Queloz,

D. (2017) Seven temperate terrestrial planets around the nearby ultracool dwarf star TRAPPIST-1. Nature 542, 456-460.

Golombek, M., Bloom, C., Wigton, N., and Warner, N. (2014) Constraints on the age of Corinto crater from mapping secondaries in Elysium planitia on Mars. LPS XXXXV, 1470.

Hartmann, W.K. (2005) Martian cratering 8: Isochron refinement and the chronology of Mars. Icarus 174, 294-320.

Hartmann, W.K., Quantin, C., Werner, S.C., and Popova, O. (2010) Do young martian ray craters have ages consistent with the crater count system?. Icarus 208:621-635.

Hawke, B.R., Blewett, D.T., Lucey, P.G., Smith, G.A., Bell, J.F., Campbell, B.A., and Robinson, M.S. (2004) The origin of lunar crater ray. Icarus 170, 1-16.

Hazael, R., Fitzmaurice, B. C., Foglia, F., Appleby-Thomas, G. J., McMillan, P. F. (2017). Bacterial survival following shock compression in the gigapascal range. Icarus, 293, 1-7.

Hazael, R., Foglia, F., Kardzhaliyska, L., Daniel, I., Meersman, F., McMillan, P. (2014) Laboratory investigation of high pressure survival in *Shewanella oneidensis* MR-1 into the Gigapascal pressure range. Front. Microbiol. 5, 612. doi:10.3389/fmicb.2014.00612.

Hazell, P.J., Beveridge, C., Groves, K., Appleby-Thomas, G. (2010) The shock compression of microorganism-loaded broths and emulsions: Experiments and simulations. Int. J. Impact Eng. 37, 433–440.

Head, J.N., Melosh, H.J., Ivanov, B.A. (2002) Martian meteorite launch: High-speed ejecta from small craters. Science 298, 1752–1756.

Hidaka, H., Sakuma, K., Nishiizumi, K., and Yoneda, S. (2017) Isotopic evidence for multi-stage cosmic-ray exposure histories of lunar meteorites: Long residence on the Moon and short transition to the Earth. The Astrophysical Journal 153:274.

Hill, D. H., Boynton, W. V., and Haag, R. A. (1991) A lunar meteorite found outside the Antarctic. Nature 352, 614-617.

Hirata, N. (2017) Spatial distribution of impact craters on Deimos. Icarus 288, 69-77.

Housen, K.R., Schmidt, R.M., and Holsapple, K.A. (1983) Crater ejecta scaling laws: Fundermental forms based on dimensional analysis. Journal of Geophysical Research 88, 2485-2499.

Holsapple, K. A. and Schmidt, R. M. (1982) On the scaling of crater dimensions: 2. Impact processes. Journal of Geophysical Research, 87, 1849-1870.

Horneck, G., Stöffler, D., Ott, S., Hornemann, U., Cockell, C. S., Moeller, R., Meyer, C., De Vera, J.-P., Fritz, J., Schade, S., and Artemieva, N. A., 2008. Microbial rock inhabitants survive hypervelocity impacts on Mars-like host planets: First phase of lithopanspermia experimentally tested. Astrobiology, 8, 17-44.

Hyodo, R., Genda, H., Charnoz, S., Pignatale, F. C. F., Rosenblatt, P., 2018. On the impact origin of Phobos and Deimos. IV. Volatile depletion. The Astrophysical Journal 860, 150(10pp).

Hyodo, R., Kurosawa, K., Genda, H., Fujita, K., and Usui, T. Extensive delivery of Martian ejecta to its moons: the gateway to a time-resolved history of Mars. in prep.

Ito, T. and Malhotra, R. (2006) Dynamical transport of asteroid fragments from the n6 resonance. Advances in Space Research 38, 817-825.

Ivanov, B. A. and Deutsch, A., 2002. The phase diagram of CaCO3 in relation to shock compression and decomposition. Physics of the Earth and Planetary Interiors, 129, 131-143.

Kawakatsu, Y., Kuramoto, K., Ogawa, N., Ikeda, H., Mimasu, Y., One, G., Sawada, H., Yoshikawa, K., Imada, Takane, Otake, Hisashi, Kusano, H., Yamada, K., Otsuki, M., and Baba, M. (2017) Mission concept of Martian Moons eXploration (MMX). 68[th] International Astronautical Congress (IAC), IAC-17-A3.3A.5.

Krijt, S., Bowling, T.J., Lyons, R.J., and Ciesla, F.J. (2017) Fast Litho-panspermia in the Habitable Zone of the TRRAPIST-1 system. The Astrophysical Journal Letters 839, L21.

Kurosawa, K. and Genda, H. (2018) Effects of Friction and Plastic Deformation in Shock-Comminuted Damaged Rocks on Impact Heating. Geophysical Research Letters, 45, 620-626.

Kurosawa, K., Okamoto, T., and Genda, H. (2018) Hydrocode modeling of the spallation process during hypervelocity impacts: Implications for the ejection of Martian meteorites. Icarus, 301, 219-234.

Lingam, M. and Loeb, A. (2017) Enhanced interplanetary pansperimia in the TRAPPIST-1 system. PNAS 114, 6689-6693.

Malin, M.C., Edgett, K., Posiolova, L., McColley, S., Noe Dobrea, E. (2006) Present impact cratering rate and the contemporary gully activity on Mars: Results of the Mars Global Surveyor extended mission. Science 314, 1557–1573.

Marvin, U. B. (1983) The discovery and initial characterization of Allan Hills 81005: The first lunar meteorite. Geophysical Research Letters 10, 775-778.

Matsumoto, M., Nishimura, T. (1998) Mersenne Twister: a 623-dimensionally equidistributed uniform pseudorandom number generator. ACM Trans. Model. Comput. Simul. 8, 3–30.

McEwen, A. S., Preblich, B.S., Turtle, E. P., Artemieva, N. A., Golombek, M. P., Hurst, M., Kirk, R. L., Burr, D. M., Christensen, P. R., 2005. The rayed crater Zunil and interpretations of small impact craters on Mars. Icarus, 176, 351–381.

McKinnon, W.B., Chapman, C.R., and Housen, K.R. (1991) Cratering of the uranian satellites. In: Bergstralh, J.T., Miner, L.D., Matthews, M.S. (Eds.), Uranus. Univ. of Arizona Press, Tucson, pp. 629–692.

Melosh, H.J. (1984) Impact ejection, spallation, and the origin of meteorites. Icarus 59, 234–260.

Melosh, H.J. (1988) The rocky road to panspermia. Nature 332, 687-688.

Melosh, H.J. (1989) Impact cratering: A geologic process. Oxford University Press, New York.

Melosh, H. J. (2011) Material Transfer from the Surface of Mars to Phobos and Deimos. Final Report: NNX10AU88G, Purdue University.

Melosh, H. J. and Ivanov, B. A. (2018) Slow impacts on strong targets bring on the heat. Geophysical Research Letters 45, 2597-2599.

Melosh, H.J. and Vickery, A.M. (1989) Impact erosion of the primordial atmosphere of Mars. Nature 338, 487-489.

Mizutani, H., Takagi, Y., and Kawakami, S. (1990) New scaling laws on impact fragmentation. Icarus 87, 307-326.

Neukum, G., Ivanov, B.A., Hartmann, W.K. (2001) Cratering records in the inner solar system in relation to the lunar reference system. Space Science Reviews 96, 55–86.

Okamoto, T., Kurosawa, K., Genda, H., and Matsui, T. Impact ejecta near the impact point observed using ultra-high-speed imaging and SPH simulations, and their comparison. Submitted.

Parkos, D., Pikus, A., Alexeenko, A., and Melosh, H.J. (2018) HCN production via Impact ejecta reentry during the late heavy bombardment. Journal of Geophysical Research: Planets 123:892-909, https://doi.org/10.1002/2017JE005393.

Patel, M., Gow, J., Paton, S., and Truscott, P., 2017. Test report on the irradiation inactivation tests results. SterLim-OU-TN15.

Patel, M., Pearson, V., Summers, D., Evans, D., Bennet, A., and Truscott, P. (2018) Sterilization Limits for Sample Return Planetary Protection Measures (SterLim)," Presentation to the Committee on the Review of Planetary Protection Requirements for Sample Return from Phobos and Deimos. ESA contract no. 4000112742/14/NL/HB.

Preblich, B. S., McEwen, A. S., and Studer, D. M., 2007. Mapping rays and secondary craters from the Martian crater Zunil. Journal of Geophysical Research E: Planets, 112, E05006.

Price, M. C., Solscheid, C., Burchell, M. J., Josse, L., Adamek, N., and Cole, M. J., 2013. Survival of yeast spores in hypervelocity impact events up to velocities of 7.4 km s$^{-1}$. Icarus, 222, 263-272.

Quintana, S. N., Schultz, P. H., and Horowitz, S. S., 2018. Experimental constraints on impact-induced winds. Icarus, 305, 91-104.

Ramsley, K. R. and Head, J. W. (2013) Mars impact ejecta in the regolith of Phobos: Bulk concentration and distribution. Planetary and Space Science 87, 115-129.

Ramsley, K. R. and Head, J. W. (2017) The Stickney crater ejecta secondary impact crater spike on Phobos: Implications for the age of Stickney and the surface of Phobos. Planetary and Space Science, 138, 7-24.

Robbins, S.J. and Hynek, B.M. (2012) A new global database of Mars impact craters >1 km: 2. Database creation, properties, and parameters. Journal of Geophysical Research 117, E05004, doi:10.1029/2011JE003966.

Schmedemann, N., Michael, G.G., Ivanov, B.A., Murray, J.B., and Neukum, G. (2014) The age of Phobos and its largest crater, Stickney. Planetary and Space Science 102, 152-163.

Schultz, P. H., 1992. Atmospheric effects on ejecta emplacement and crater formation on Venus from Magellan. Journal of Geophysical Research, 97, 16,183-16,248.

Schultz, P. H. and Quintana, S. N., 2017. Impact-generated winds on Mars. 292, 86-101.

Schmidt, R.M. and Housen, K.R. (1987) Some recent advances in the scaling of impact and explosion cratering. International Journal of Impact Engineering 5:543-560.

Shoemaker, E.M. (1962) Interpretation of lunar craters. in Physics and Astronomy of the Moon, edited by Z. Kopal, pp. 283–359, Academic, San Diego,Calif.

Shuvalov, V. (2009) Atmospheric erosion induced by oblique impacts. Meteoritics & Planetary Science, 44, 1095-1105.

Strauss, J. H. and Sinsheimer, R. L. (1963) Purification and properties of bacteriophage MS2 and of its ribonucleic acid. Journal of Molecular Biology, 7, 43-54.

Sugita, S. and Schultz, P. H. (2002) Initiation of run-out flows on Venus by oblique impacts. Icarus, 155, 265-284.

Summers, D. (2017) Evaluation of the level of assurance that no unsterilized martian material naturally transported to Phobos (and Deimos) is accessible to a Phobos (and Deimos) sample return mission. SterLim-Ph2-TAS-TN21.

Tanaka, H., Inaba, S., and Nakazawa, K. (1996) Steady-state size distribution for the self-similar collision cascade. Icarus 123, 450-455.

Thomas, P. C. (1998) Ejecta emplacement on the Martian Satellites. Icarus 131, 78-106.

Thomas, P. C., Veverka, J., Sullivan, R., Simonelli, D. P., Malin, M. C., Caplinger, M., Hartmann, W. K., and James, P. B. (2000) Phobos: Regolith and ejecta blocks investigated with Mars Orbiter Camera images. Journal of Geophysical Research, 105, 15,091-15,106.

Tornabene, L.L., Moersch, J.E., McSween, H.Y., McEwen, A.S., Piatek, J.L., Milam, K.A., and Christensen, P.R. (2006) Identification of large (2-10 km) rayed craters on Mars in THEMIS thermal infrared images: Implications for possible Martian meteorite source regions. Journal of Geophysical Research 111, E10006, doi:10.1029/2005JE001600.

van Duin, J. and Tsareva, N. (2006) Single-stranded RNA phages. Chapter 15 (pp. 175-196) in: Calendar RL (ed.), The Bacteriophages (Second Edition). Oxford University Press.

Vickery, A.M. and Melosh, H.J. (1987) The large crater origin for SNC meteorites. Science 237, 738–743.

Werner, S.C., Ody, A., and Poulet, F. (2014) The source crater of Martian Shergottite meteorites. Science 343, 1343-1346.

Wiens, R.C. and Pepin, R.O. (1988) Laboratory shock emplacement of noble gases, nitrogen, and carbon dioxide into basalt, and implications for trapped gases in shergottite EETA 79001. Geochimica et Cosmochimica Acta 52, 295-307.

Willis, M.J., Ahrens, T.J., Bertani, L.E., Nash, C.Z., 2006. Bugbuster-survivability of living bacteria upon shock compression. Earth Planet. Sci. Lett. 247, 185–196. http://dx.doi.org/10.1016/j.epsl.2006.03.054.

Zahnle, K., Schenk, P., Levison, H. and Dones, L. (2003) Cratering rates in the outer solar system. *Icarus*, *163*(2), 263 – 289, doi:http://dx.doi.org/10.1016/S0019-1035(03)00048-4.